# Dark Brain Energy: Toward an Integrative Model of Spontaneous Slow Oscillations

Zhu-Qing Gong[1,2,3] and Xi-Nian Zuo[1,2,3,4,5,6,7,8*]

[1]State Key Laboratory of Cognitive Neuroscience and Learning, Faculty of Psychology, Beijing Normal University, Xinjiekouwai Street 19, Haidian District, Beijing, 100875, China.

[2]Department of Psychology, University of Chinese Academy of Sciences, No 19 Yuquan Road, Shijingshan District, Beijing, 100049, China.

[3]Key Laboratory of Behavioural Sciences, Institute of Psychology, Chinese Academy of Sciences, No 16 Lincui Road, Chaoyang District, Beijing, 100101, China.

[4]National Basic Science Data Center, No 2 Dongsheng South Road, Haidian District, Beijing, 100190, China.

[5]Key Laboratory of Brain and Education, School of Education Sciences, Nanning Normal University, No 175 Mingxiu East Road, Mingxiu District, Nanning, Guangxi, 530001, China.

[6]Research Center for Lifespan Development of Mind and Brain, Institute of Psychology, Chinese Academy of Sciences, No 16 Lincui Road, Chaoyang District, Beijing, 100101, China.

[7]Developmental Population Neuroscience Research Center, IDG/McGovern Institute for Brain Research, Beijing Normal University, No 19 Xinjiekouwai Street, Haidian District, Beijing, 100875, China.

[8]Engineering Center for Population Neuroimaging and Intellectual Technology, Nanning Normal University, No. 175, Mingxiu East Road, Nanning 530001, China

*Correspondence: Xi-Nian Zuo

Beijing Normal University, Xinjiekouwai Street 19, Haidian District, Beijing, 100875, China.

E-mail address: xinian.zuo@bnu.edu.cn, zuoxn@psych.ac.cn, zuoxn@nnnu.edu.cn.

**Abstract**

Neural oscillations facilitate the functioning of the human brain in spatial and temporal dimensions at various frequencies. These oscillations feature a universal frequency architecture that is governed by brain anatomy, ensuring frequency specificity remains invariant across different measurement techniques. Initial magnetic resonance imaging (MRI) methodology constrained functional MRI (fMRI) investigations to a singular frequency range, thereby neglecting the frequency characteristics inherent in blood oxygen level-dependent oscillations. With advancements in MRI technology, it has become feasible to decode intricate brain activities via multi-band frequency analysis (MBFA). During the past decade, the utilization of MBFA in fMRI studies has surged, unveiling frequency-dependent characteristics of spontaneous slow oscillations (SSOs) believed to base dark energy in the brain. There remains a dearth of conclusive insights and hypotheses pertaining to the properties and functionalities of SSOs in distinct bands. We surveyed the SSO MBFA studies during the past 15 years to delineate the attributes of SSOs and enlighten their correlated functions. We further proposed a model to elucidate the hierarchical organization of multi-band SSOs by integrating their function, aimed at bridging theoretical gaps and guiding future MBFA research endeavors.

**Keywords**: spontaneous brain activity, slow oscillation, neuroimaging, multi-band frequency analysis, network neuroscience, connectome, dark brain energy, resting-state fMRI

## Introduction

Brain rhythms, from the whispers of a single neuron to the orchestral crescendo of billions, hold the key to unlocking the secrets of our cognitive universe. These patterns of neural activity or fluctuations oscillate as the heartbeat of our cognitive world. Neural oscillations in the brain cover a very wide range, from ultraslow oscillations with frequencies lower than 0.01 Hz to very high oscillations surpassing 1000 Hz [1, 2]. The first observation and investigation of these oscillations in the human brain date to the early 20th century, when Berger discovered alpha and beta waves using electroencephalogram (EEG) devices [3]. Since then, an increasing number of neuroscientists have examined the oscillations in the human brain and questioned the physiological mechanisms underlying these oscillations and their relationship to mental states and behaviors. Multiple techniques have been adopted to directly or indirectly elucidate the mechanism of brain oscillations at different levels. Single-unit recording directly measures the electrophysiological responses of single neurons at the microscopic level. EEG, magnetoencephalography (MEG) and functional magnetic resonance imaging (fMRI) measure neural oscillations at indirect and macroscopic levels, reflecting the population behavior of neurons. For nearly a century, electrophysiological studies have gradually outlined the functional characteristics of, and the relationships between, different frequency bands specific to states and functions. Specifically, low frequencies dominate long-range connections, indicating top-down processing, multimodal integration, and coupling between remote areas, whereas high frequencies are active in primary sensory and motor regions and prevalent in local integration [4-6]. This frequency character is supported by the brain's anatomy, wherein primary brain regions are linked by highly myelinated thick fibers that transfer information quickly, while higher-order brain regions are connected by less myelinated thin fibers that transfer information slowly [7,8]. The brain's anatomical organization ensures that its frequency specificity is independent of measuring techniques. Therefore, from a holistic perspective, neural activity measured in different scales by different technologies should exhibit similar frequency characteristics.

This similarity in frequency characteristics is supported by the properties of complex systems. These neural activities organized by the brain as a complex system, can be measured at various scales or levels using different methods. The fractal characteristics of the complex system indicate measurements of the same system at different scales exhibit self-similarity [9,10]. Specifically, from a macroscopic perspective, neural activity measured by different instruments may also exhibit comparable frequency characteristics, with both high- and low-frequency components showing similar properties. This extends to spatial dimensions as well. For instance, neural activity observed at different scales within the visual cortex is consistently linked to visual information processing. Besides the similarities among different measuring scales, distinct laws or principles emerge at each scale. While the frequency characteristics of individual neurons can be studied, activation patterns of neural activities at higher population scales cannot be directly inferred from basic scales [11]. Additionally, it is impractical to measure each neuron individually given their vast scale. Instead, these patterns must be understood within the context of the scale at which they are observed. Thus, investigating the frequency and spatial characteristics of neural activity at each measurement scale offers equally important insights to understand how the brain functions.

The electrophysiological techniques have millisecond-level sampling rates, which allows capturing the frequency characteristics of fast neural oscillations related to various cognitive functions including memory formation, emotion processing, attention, action, and higher cognitive performance of insight and even spiritual experiences. However, spatial distribution of the neural oscillations is poorly defined due to the low spatial resolution. Human neuroimaging, with fMRI, has revolutionized the research landscape as one of the most popular techniques for detecting brain's responses to various cognitive stimuli in vivo at very high spatial resolutions. It measures the blood oxygen level-dependent (BOLD) signal, which is based on local blood flow changes due to the controlled stimuli. Previous studies have revealed that the additional energy required for such task-oriented responses is extremely small, leaving most of the ongoing amount of energy that the brain costs for unknown functions. This 'dark energy' in the brain has been considered as the primary energy consumption of the brain to support the daily life of spontaneous neural oscillations in maintaining an intrinsic functional architecture in the brain [12]. We note that fMRI compresses brain oscillations into a relatively low-frequency landscape, providing a unique view of the slow waves of spontaneous neural oscillations that have rarely been detected using EEG and MEG in previous studies. Although this measure is indirect, its neural origin has been well-characterized by numerous studies [13]. Neural activities induce changes in the volume of blood flow, leading to changes in the concentrations of oxyhemoglobin and deoxyhemoglobin [14, 15]. This concentration change is reflected in the BOLD signal; therefore, the BOLD signal is considered a physiological indicator of neural activity [16]. In addition, BOLD signals directly reflect neural activity measured by single- or multi-unit recordings, as well as local field potentials (LFPs) [17-19], among which LFPs can be best estimate of BOLD signals than multi-unit recording, indicating that BOLD signals reflect a population level of neural activity [20].

Since the frequency pattern conveys information in neural activity, the frequency character in the BOLD signal should not be omitted. However, in most fMRI studies, researchers primarily focused on the spatial characters of neural activity, largely overlooking the frequency characteristics of BOLD oscillations. The BOLD signal has been studied as an entire band of oscillations rather than as a compound signal that must be further divided into oscillations with different physiological mechanisms. Specifically, the researchers preferred to filter the original BOLD signal into a single band, which was assumed to contain most of the neural information. In their seminal work, Dr. Biswal and colleagues developed the resting-state fMRI (rfMRI) protocol to detect spontaneous brain activity, a powerful and popular method of investigating the spontaneous neural oscillation at slow frequencies [21,22]. They discovered temporal correlations among the time courses of spontaneous slow oscillations (SSOs) (<0.1 Hz) in the motor areas of the resting-state human brain. It generated the first map of the human resting-state motor network through investigations into spontaneous brain activities [23]. In the majority of subsequent rfMRI studies, the frequency range chosen for analysis follows this work and ranges from approximately 0.01 Hz to 0.1 Hz, for several reasons. First, frequencies below 0.01 Hz are considered system noise caused by scanner drift and coil interference. Second, respiratory fluctuations, which peak at approximately 0.3 Hz, and pulse waves, which peak at 1 Hz, may interfere with neuronal signals [24, 25]. Therefore, filtering the BOLD signal to this frequency range is often used as a preprocessing denoising technique. Finally, as depicted in the seminal paper by Dr. Biswal [21], the power spectrum of the scanning signal is visually dominated by low frequencies less than 0.1 Hz while nearly flattening to zero at frequencies greater than 0.1 Hz. This cutoff frequency (0.01 to 0.1 Hz) has become a standard in rfMRI studies since then, leading to substantial progress in mapping the intrinsic connectivity networks and understanding their roles related to the brain's dark energy [26].

Despite these advances, this setting remains primarily empirical and lacks a criterion supported by scientific theory. Different studies employed different frequency bands; for example, one study focused on 0.009 Hz to 0.08 Hz [27], while another analyzed signals from 0.06 Hz to 0.11 Hz [28]. The choice of different frequencies may pose a potential problem when comparing results from different studies (e.g., meta-analyses). Critically, BOLD signals outside the range of 0.01 Hz to 0.1 Hz may also contain biologically plausible information on neural oscillations, so filtering out signals outside that range results in a loss of the full landscape of neural information. As shown in the rfMRI data from a healthy young adult in the Human Connectome Project (HCP) presented in Figure 1 [29, 30], the human amygdala exhibits an amplitude profile equally distributed across all detectable frequencies by using both 3T and 7T scanners, while the canonical cortical networks show power-law amplitude profiles. This indicates that, with the improvement of MRI instrument performance and scanning sequence, BOLD signals are distributed over the entire detectable frequency domain rather than being concentrated only around 0.01 Hz to 0.1 Hz. Therefore, using other noise reduction methods, such as those based on independent component analysis to remove noise components, rather than directly filtering out all signals outside a certain frequency range, is more appropriate. Ultraslow (less than 0.01 Hz) BOLD signals exhibit a similar tissue-specific spatial pattern to traditional SSOs (0.01 Hz to 0.1 Hz) and a strong positive correlation with BOLD signals during cognitive task performance, indicating that SSOs less than 0.01 Hz represent spontaneous brain physiology rather than scanner noise [31]. SSOs can also be observed using noise-free methodology by separating physiological noise components from the default mode network component at frequencies greater than 0.25 Hz [32]. Overall, full-frequency SSO BOLD signals deserve further investigation.

A single-band analysis of the compound signal might conceal or mix information carried by individual oscillations that constitute the frequency range and have specific functions. Multi-band analysis outperforms single-band analysis in disease classification [33]. The energy contrast between frequency bands can reliably characterize different brain states [34]. Both remote [35,36] and local [37,38] measurements of BOLD signals have revealed frequency-dependent characteristics of SSOs in the human brain. These findings are consistent with both EEG [39, 40] and single-neuron recording [41] studies, which show that different brain activities preferentially function at different frequencies. Figure 1 depicts a symphony in the brain operated by the neuronal orchestra based on the large-scale intrinsic connectivity networks. Although this offers a nice understanding of the brain's intrinsic functional architecture, we need further information of the different playing neuronal musicians (i.e., SSOs at different frequency bands) to unlock mechanisms behind the brain rhythms. Therefore, multi-band analysis is more informative and suitable than single-band approaches for decoding brain function through BOLD signals. The amplitude of low-frequency fluctuations (ALFF) of the thalamus is higher in older adults than in younger adults for SSOs from 0.01 Hz to 0.1 Hz. However, dividing this standard frequency band into multiple subbands revealed that age-related ALFF changes were dominated by SSOs with frequencies less than 0.027 Hz, while SSOs with frequencies from 0.198 Hz to 0.250 Hz showed the opposite trend [42]. This example highlights the necessity of decomposing BOLD signals into multiple SSOs across different frequency bands to further understand their functions and underlying mechanisms. In recent years, an increasing number of studies, especially clinical studies, have accumulated evidence that different frequency bands of BOLD SSOs process distinct activity characteristics. However, these studies remain at the level of phenomenological description of observational data. At present, there is a lack of a unified theory to explain the phenomena observed in different multi-band fMRI studies. The overall aims of the present review are to 1) introduce a neural frequency architecture as theoretical guidance for decomposing BOLD SSO signals into multiple frequency bands; 2) systematically survey SSO studies using fMRI with multi-band frequency analysis (MBFA) published in the past 15 years to summarize the characteristics of SSOs and extract their associated functions; and 3) propose a theoretical model to describe the hierarchical organization of multi-band SSOs, aiming to link data and theory for understanding the brain's 'dark energy' to foster future SSO research.

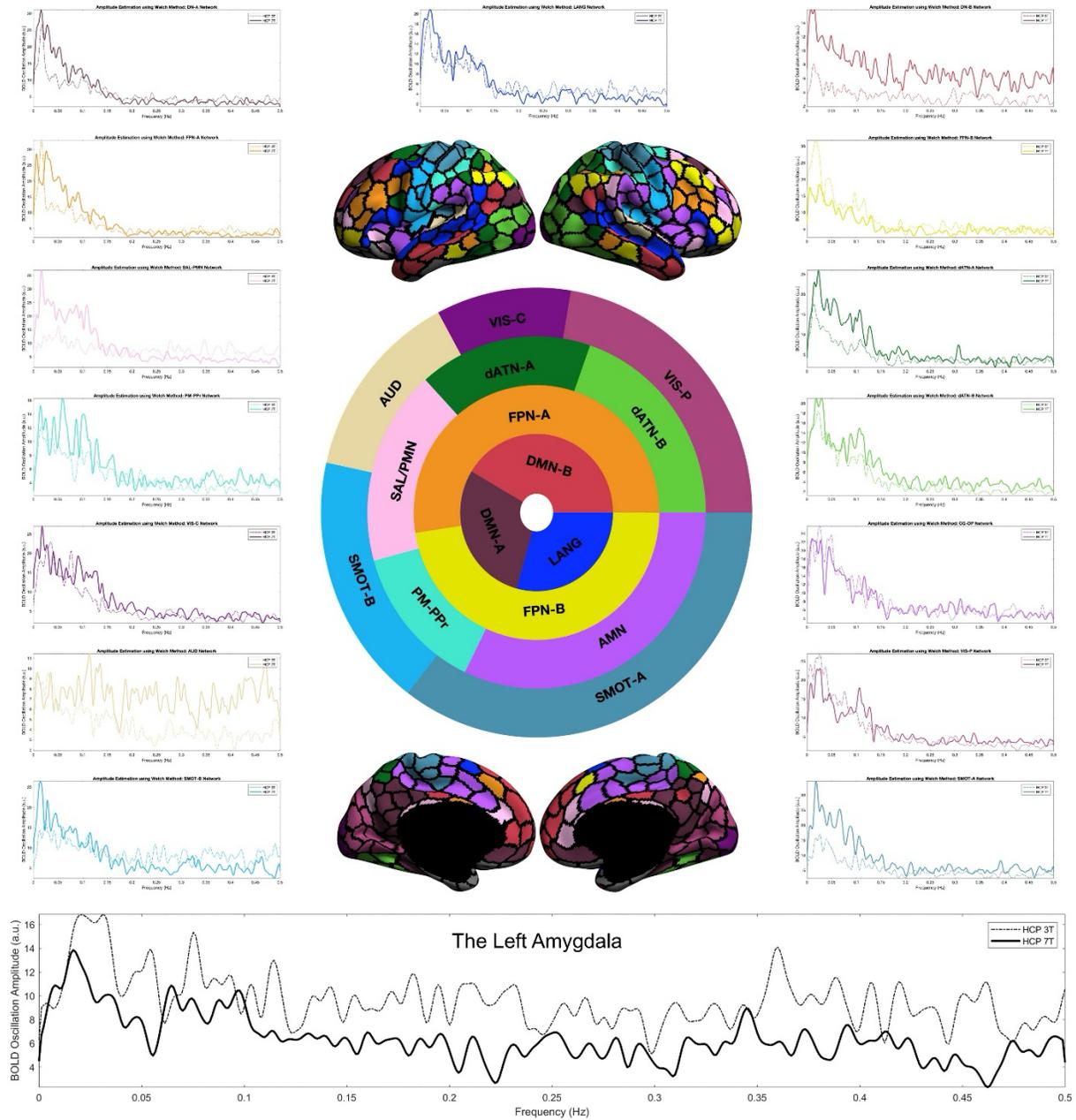

**Figure 1. Symphony of *spontaneous slow oscillations (SSOs)* in the human brain**. *The whole brain of a healthy young adult volunteer participating in the Human Connectome Project (HCP subid 100160) was scanned with the BOLD rfMRI protocol using both 3.0 Tesla (3 T) and 7.0 Tesla (7 T) magnets. Network-specific signals were obtained by averaging the BOLD signals for all the vertices within each of the fifteen canonical cortical networks defined in a previous study: Somatomotor-A (SMOT-A), Somatomotor-B (SMOT-B), Premotor-Posterior Parietal Rostral (PM-PPr), Action-Mode Network (AMN), Salience/Parietal Memory Network (SAL/PMN), Dorsal Attention-A (dATN-A), Dorsal Attention-B (dATN-B), Frontoparietal Network-A (FPN-A), Frontoparietal Network-B (FPN-B), Default-Mode Network-A (DMN-A), Default-Mode Network-B (DMN-B), Language (LANG), Visual Central (VIS-C), Visual Peripheral (VIS-P), and Auditory (AUD)[58,59]. All the networks are rendered onto the human cerebral cortex with 400 parcellating units [57,60,61]. For each network, the amplitudes of its SSO signal are plotted against frequencies obtained using both 3 T (dashed lines) and 7 T magnets (solid lines). This visualization was also performed for the SSO of the left amygdala, as shown in the bottom panel. The BOLD oscillation amplitude of the left amygdala is almost uniformly distributed across the detectable frequencies, while the SSO amplitude distributions in the cortical networks follow the power law. Beyond the differences in the distributions of amplitudes, the SSOs in BOLD signals are detectable throughout the entire measurable frequency range, which is a reproducible pattern for both 3 T and 7 T magnets.*

## A theory-based Slow Oscillation Taxonomy

In 2003, Penttonen and Buzsáki first proposed the natural logarithm linear law (N3L) to describe the frequency architecture of neural oscillations in the mammalian brain [43]. According to the N3L, neural oscillations form a linear hierarchy of multiple frequency bands when transferred to the natural logarithmic scale. Specifically, neural oscillations show a linear progression on the natural logarithmic scale, in which the center frequencies fall on the integers of the logarithmic axis such that the distance between the centers of adjacent oscillations have a constant value of 1. On the natural number axis, neural oscillations exhibit a geometric progression where the center frequencies between adjacent oscillations have a constant ratio, which approximately equals $e$, the base number for the natural logarithm. These oscillations predicted by the N3L have been observed in various independent experiments with little overlap of frequency ranges, which indicates that these bands have independent mechanisms. The N3L not only describes the quantitative relationship among observed oscillations but also predicts oscillations along the frequency axis that have not yet been recorded or investigated with currently existing technologies.

Although the N3L was first calculated and estimated based on EEG signals, it reveals a unifying characteristic of the physiological oscillation system that can be measured with different techniques and provides a uniform taxonomy for these oscillations. Both behavioral slow fluctuations and physiological rhythms have been fitted using this hierarchical relationship [44,45]. For BOLD signals, the currently detectable frequency ranges roughly cover the slow oscillation class predicted by the N3L theory, namely, slow-$n$ ($n$=1, ... ,6), as shown in Figure 2. The N3L is a theory-based taxonomy for classifying BOLD signals that provides a frequency architecture to integrate oscillations measured directly (neuron) or indirectly (behavior) at different levels. There is a noteworthy factor when utilizing the N3L to decompose the BOLD signals. Oscillations are continuous in theory, while the signals acquired using brain imaging equipment are discrete in practice. According to the sampling theory proposed by Nyquist and Shannon, the sampling frequency and duration determine both the upper and lower frequency boundaries that can be measured and constructed. Thus, BOLD signals acquired using fMRI with different parameters should have different detectable frequency ranges. In addition, the sampling parameters also determine the frequency resolution. In practice, these parameters lead to different numbers of decomposable frequency bands and differences in the precise boundaries of each band. Therefore, researchers must use the imaging parameters to calculate the precise boundaries of decomposable frequency bands, rather than referring to the bands from a previously published paper [46]. The main reason for this situation is that a professional tool for BOLD signal decomposition is lacking for those who are not familiar with the N3L applications. We thus developed a graphical interface toolbox, DREAM [44], based on the N3L theory to decompose both neural and physiological oscillations (e.g., head movement) more precisely.

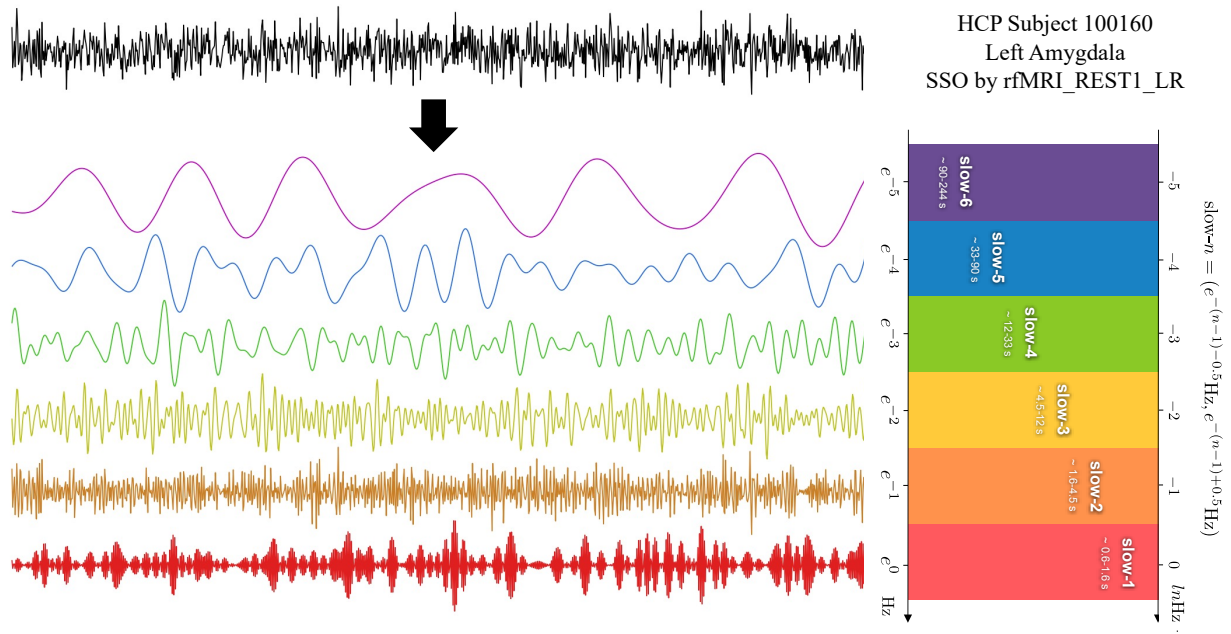

**Figure 2.** *Classification of slow oscillations based on the natural logarithm linear law*. As depicted, each frequency band (i.e., the slow-n, n=1,2,3,4,5,6) is centred at the integer point -(n-1) in the natural log axis with a width of 1. On the original frequency axis, the centre of each frequency band is located at $e^{-(n-1)}$. The right panel shows the slow oscillations from the mean BOLD rfMRI signals of the left amygdala (i.e., REST1_LR scan from the HCP100160 volunteer).

In this review, we define BOLD oscillations as distinct frequency bands derived from the full-spectrum BOLD signal through frequency decomposition. Specifically, the commonly defined resting-state frequency range (0.01 to 0.1 Hz) represents a sub-band of the full-spectrum BOLD signal, based on an empirical division. Similar to electrophysiological recordings, the full-spectrum BOLD signal exhibits *1/f* characteristics, or avalanche dynamics, where the power spectral density of the signal inversely correlates with frequency [48,49]. This reflects the complexity of neural activity and suggests that low-frequency oscillations dominate high-frequency activity [50]. Notably, the scale-free nature of neural activity does not contradict its scale-specific characteristics; instead, both aspects coexist within the dynamics of the brain [51,52]. Each BOLD frequency band can be viewed as a segment of the original full-spectrum *1/f* signal, preserving the scale-free self-similarity of the entire signal. However, a distinction must be made between BOLD oscillations and neural oscillations as measured in electrophysiological studies. The latter typically emphasize signal periodicity [53,54], whereas BOLD oscillations focus on reflecting the functional organization across frequency bands, often without addressing the periodicity within individual bands. In this review, we summarize the functional characteristics of each N3L-based BOLD frequency band and propose theoretical hypotheses concerning the functional hierarchy and interactions between these bands, to integrate the scale-free and scale-specific properties of BOLD signals from a macro-level perspective on the full spectrum.

## A Bibliometric Analysis of N3L-based SSOs Studies

In 2010 [46], inspired by an MBFA study on intraindividual variability of behavioral performance [44], Zuo and colleagues used the N3L to examine the multi-band frequency characteristics of rfMRI signals. They mapped the spatial distributions of SSO amplitudes for the slow-2, slow-3, slow-4 and slow-5 bands, as well as their test-retest reliability. Of particular interest, mapping the differences in the SSO amplitudes among these frequency bands revealed specificity of the slow-4 SSO amplitude for the human basal ganglia. This finding represents one of the earliest pieces of evidence that neuroimaging findings may be frequency-specific and may benefit from theory-driven MBFA. We provide a bibliometric analysis of this work to assess the tangible benefits of using the N3L-based decomposition method to identify SSOs [46].

Our aim in conducting this analysis is to not only provide a comprehensive review of the N3L-based MBFA in fMRI research but also inspire the subsequent generation of hypotheses and knowledge on the frequency architecture of human brain function. As of January 2025, the paper has been cited nearly 1,500 times according to Google Scholar[1]. After excluding repeated citations and those without full-text support, we confirmed 1,490 studies that cited the paper. A total of 204 of these studies adopted MBFA to investigate human brain function using fMRI. We subsequently conducted a literature search on *Web of Science* with *'multiple frequency'* and *'fMRI'* as the keywords to retrieve studies published from January 2010 to January 2025 and to avoid omitting studies by only focusing on citations of this single paper. Among the 1034 retrieved papers, only 20 studies were missing from the bibliometric analysis: 5 MBFA studies using the N3L method and 15 MBFA study adopting other frequency decomposition methods. We thus added the omitted 20 studies to the bibliometric analysis and reviewed all 204 papers on fMRI signal MBFA published over the past decade.

The distributions of publication years and research fields of these studies are depicted in Figure 3. The numbers of publications on both basic and clinical MBFA research increased with publication year, with more than three-quarters of the MBFA studies being clinical trials. This result aligns with the overall impression that MBFA of human brain SSOs offers novel, reliable, and potentially valid neuroimaging methods for examining SSOs altered by various clinical conditions [55,56]. However, the fundamental theory regarding the specific roles of SSOs and the mechanisms of their interactions across multiple frequency bands remains largely unknown. Therefore, interpreting MBFA clinical findings is challenging. Meanwhile, basic MBFA research appears to have progressed more slowly, indicating the challenges in parsing the biological validity of these SSOs and the need for theoretical research progress. We used N3L theory as a guideline for surveying MBFA studies utilizing both N3L-based and other frequency decomposition methods. First, we summarized the spatial patterns and associated functions of each N3L-derived frequency band from slow-6 to slow-1 by mapping the reporting cortical coordinates from these studies onto the parcellation areas in the 400 units of the 15 canonical intrinsic brain networks as in Figure 1. Then, we proposed a hierarchical model of these slow-band SSOs to describe their organization and functional relationships. Using this approach, we attempt to identify several challenges and opportunities of MBFA in the discovery of biomarkers to foster clinically useful applications and their future translations.

---

[1] https://scholar.google.com/citations?view_op=view_citation&citation_for_view=a3-gVGMAAAAJ:d1gkVwhDpl0C

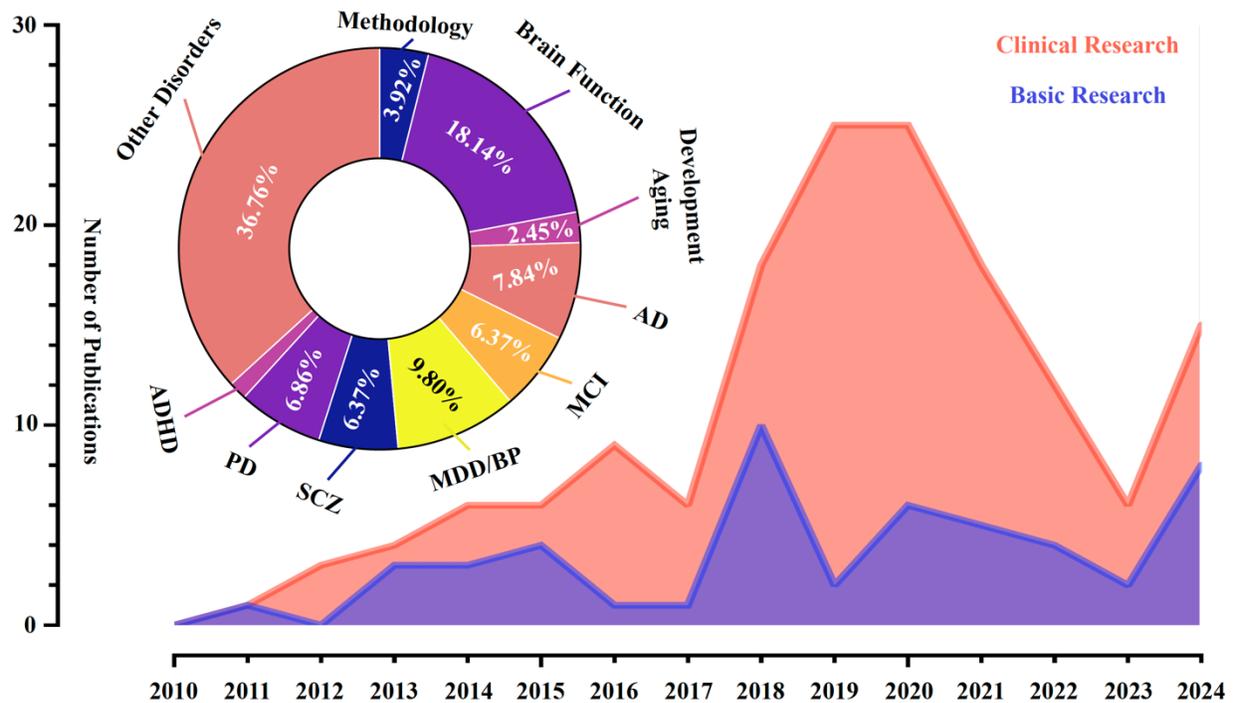

*Figure 3. Distributions of the publication years and research fields. The numbers of publications on both basic and clinical research increased with publication year, and more than three-quarters of the studies were clinical trials.*

## Characterizing the N3L-derived SSOs

We collected coordinates of all the locations reported in the papers from the bibliometric analyses and converted them into the standard brain space. We rendered the central coordinate of each reported region into a single parcel from the local-global parcellation with 400 parcels of the 15 resting-state networks to summarize the network-level associations of these locations [57-61]. This analysis was performed for each of the 6 slow bands. As shown in Figure 4, slow-2 to slow-6 SSOs were associated with network-wise functional parcellation, while slow-1 SSOs did not show the same association due to the limitation of the rfMRI sample rate in previous studies. Band-specific characteristics are documented in the subsections below according to the empirical network-level summaries.

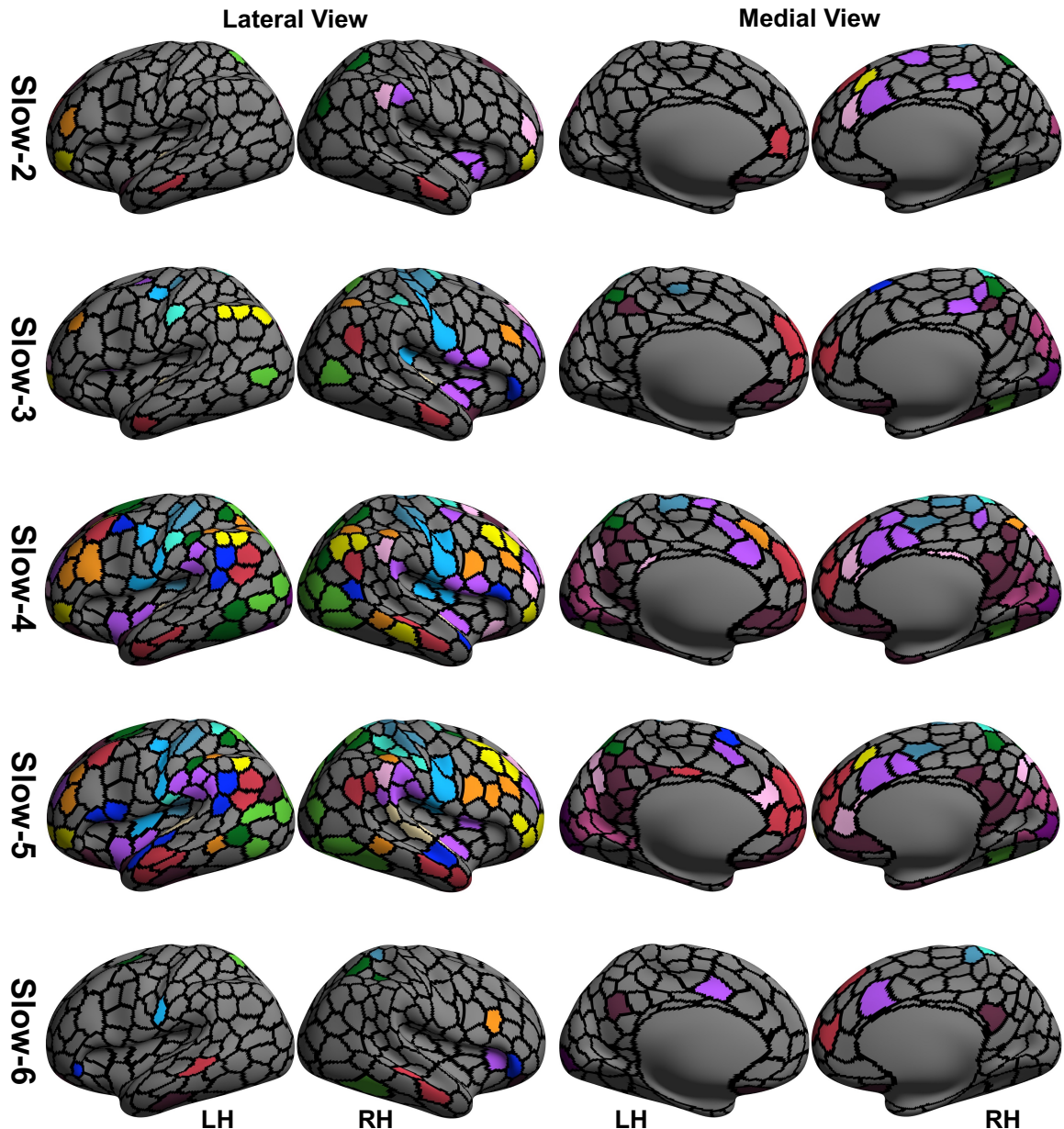

*Figure 4. Cortical areas with active SSO profiles in nonclinical or clinical conditions for each slow band. The colors indicate the networks to which these regions belong as denoted in Figure 1. The black lines indicate the boundaries between the parcellating areas.*

**Slow-6: Long-range integration and top-down modulation**

The slow-6 SSOs oscillate between approximately 0.0041 and 0.0111 Hz. In early studies, the main component of signals below 0.01 Hz was considered a slow drift induced by scanner instability [62]. Therefore, slow-6 SSOs were often filtered out as noise components in most rfMRI studies and thus have rarely been studied. However, are slow-6 SSOs only dominated by scanner noise? One study systematically explored the BOLD signal characteristics of the ultraslow-frequency band (0 Hz to 0.01 Hz) and provided a negative answer to this question [31]. First, the distribution of the signal was tissue-specific, similar to the traditional rfMRI interval (0.01 to 0.1 Hz). Second, the effect of scanning parameters on the signal was similar to that of task-induced changes in BOLD signals but differed from those of phantom experiments. Third, the baseline magnitude of the signal strongly correlated with task-induced changes in BOLD signals. Overall, the study revealed that the major component of slow-6 SSOs is spontaneous physiological fluctuations rather than scanner noise or physiological noise, such as cardiopulmonary and respiratory noise. Therefore, overlooking this frequency range would neglect the potential neural information conveyed through ultraslow-frequency oscillations.

To date, a few studies using MBFA have delineated the characteristics of slow-6 SSOs. The amplitude distribution in slow-6 showed a similar pattern to that of slow-5, suggesting functional similarities between the two bands. Specifically, compared to higher-frequency oscillations, especially those higher than slow-3, the amplitude ranking is higher in slow-6 for the lateral part of frontoparietal network-A (FPN-A), the ventral part of dorsal attention-B (dATN-B), and the temporal regions of default mode network-B (DMN-B) [47]. The network properties showed that SSOs in slow-6 contribute more to functional integration. In the resting state, slow-6 oscillations exhibited more global connectivity across the cortex than those in the traditional frequency interval [31]. This observation is consistent with previous EEG findings that long-range integration occurs in low-frequency ranges [6,63]. The long-range interactions observed in EEG studies are usually dominant without sensory inputs, reflecting top-down, internally driven activities. Do these global connections in slow-6 BOLD oscillations represent top-down integration as well? A positive answer to this question is supported by a study using graph theory [36]. Most intrinsic connectivity networks, such as the default mode, language, somatomotor, salience, and visual networks, contribute more to global efficiency in the lower frequency band (0.009 to 0.012 Hz) than in the higher frequency band (0.03 to 0.08 Hz). Therefore, most brain networks play an integrative role through global connections in the low-frequency band. In addition, clinical findings have provided parallel evidence. Children with attention deficit hyperactivity disorder (ADHD) exhibited the largest differences in the regional distribution of functional homogeneity (ReHo) in slow-6 compared to typically developing children [64]. The altered ReHo metrics were mainly distributed in brain regions across the default mode, language, frontoparietal and dorsal attention networks, which are associated with top-down internal activities. Inattentiveness is a major symptom of this disorder and may be related to disturbances in these top-down functions in slow-6 SSOs.

Recent research also suggested that the default mode network is detectable in slow-6 [65] with a highly similar connection pattern to the network detected in slow-5, although the connection strengths are weaker. Interestingly, the default mode network nodes displayed different levels of degree centrality across different frequency bands. This result may indicate functional segregation between bands in the frequency domain. Specifically, the right superior frontal gyrus shows the highest degree centrality in slow-6, indicating it as a core hub region of the default mode network in this frequency band. Additionally, aberrant functioning of this area in slow-6 has been reported in patients with various mental disorders, further suggesting this region's hub role in slow-6 [64,66]. In summary, the slow-6 band plays an important role in global information integration and top-down functions such as attention control, with the right superior frontal gyrus serving as an oscillating hub.

**Slow-5: Default core, consciousness and memory**

The slow-5 SSOs oscillate between approximately 0.0111 Hz and 0.0301 Hz. Along with slow-4, this band is part of the traditional frequency interval of the rfMRI signal. The results of most rfMRI studies reflect the combined characteristics of these two bands, obscuring some of the unique features of each band. Compared to slow-4 and higher frequency bands, slow-5 exhibits more remote connections, similar to those of slow-6. Most networks exhibit stronger connectivity and denser between-network connections. Conversely, the default mode network, cingulo-opercular network, and ventral attention network exhibited stronger within-network connections [35]. This wide range of remote connections is analogous to the coarser synchronization in low-frequency EEG signals, reflecting long-range coupling between neurons in distant areas [5]. This network characteristic implies that information is integrated among different cortical networks in slow-5, with complex cognitive functions requiring the participation of different network regions dominated by SSOs in this frequency band. If this hypothesis is correct, disconnections of these pathways would cause failures of cognitive functions requiring the cooperation of multiple modalities. Evidence from studies of major depressive disorder (MDD) supports this hypothesis. Patients with MDD typically suffer impairments in many functional domains, including emotional processing, affective cognition, cognitive control, and reward processing. These functions involve top-down regulation, integrating internal activities with external inputs and thus requiring the collaboration of multiple network regions. In slow-5, patients with MDD showed large-scale disconnections among brain network regions [67]. Specifically, in treatment-sensitive patients with MDD, disengaged brain regions were primarily detected in the frontoparietal network, while no significant disconnections were observed in slow-4. However, in treatment-resistant patients with MDD, many more network disconnections were detected across the cortex, including the orbital prefrontal, limbic, lateral parietal, lateral temporal, and medial/inferior occipital regions. These disconnections may reflect that the pathways supporting the long-range coupling of brain regions are blocked or disrupted in patients with MDD, leading to the functional failure of information integration. Connectivity strength among regions such as the posterior cingulate cortex, presupplementary motor area, insula, amygdala, and hippocampus was also reported to increase in patients with MDD. Conversely, decreases in connectivity are associated with anhedonia, while increases in connectivity are associated with the severity of depressive symptoms [68]. Alterations in slow-5 functional connectivity (FC) were also observed in patients with other mental or neural disorders, such as schizophrenia [69,70], bipolar disorder [71], mild cognitive impairment [72], Alzheimer's disease [73], and Parkinson's disease [74,75]. However, the mechanisms underlying these changes in FC in patients with these disorders remain unknown and require further investigation.

Although most networks exhibit global functional integration in slow-5, the default mode network appears to segregate from other networks and integrate information within itself at this frequency [35]. The default mode network exhibited the highest power in the slow-5 band, indicating its peak activity in this frequency range [38]. The default mode network areas displayed the strongest connectivity strength and the highest connectivity density among each other, retaining a relatively high degree of centrality in slow-5 [38,65]. Its posterior-medial core region, which encompasses most brain areas of the default mode network-A (DMN-A), including the precuneus and posterior cingulate cortex, exhibited the highest power spectrum in slow-5, along with higher ReHo, indicating that the functions of these two regions are relatively homogeneous in slow-5 [37,76,77]. The precuneus is suggested to be crucial for maintaining consciousness and performing conscious information processing and is involved in self-consciousness, episodic memory, and visuospatial-related attention shifts [78-82]. The distribution of the high slow-5 power spectrum within these two regions at rest is consistent with previous findings that they exhibited the highest resting-state cerebral glucose metabolism, which showed the greatest reduction under anesthesia [83]. Therefore, these structures play an important role in maintaining consciousness. As the default mode network shows greater within-network integration and fewer between-network connections in slow-5 SSOs, we postulate that these slow-5 SSOs activate the default core regions, which are primarily involved in consciousness and internal information processing.

Another functional role of slow-5 SSOs is associated with memory. Among the low frequencies sampled, the parietal memory network was only present in the slow-5 band [84]. Additionally, the slow-5 band exhibited the best performance in classifying the early stage from the late stage of mild cognitive impairment [72]. Widespread areas exhibiting abnormal slow-5 SSOs were observed in patients with cognitive memory deficits [73,85-91]. These alterations, particularly in the functions of the precuneus, are greater in slow-5 than in other higher frequencies [85,86,88,92]. This region is a core area overlapping the DMN-A and the parietal memory network. The abnormal slow-5 SSOs in the precuneus may partly account for the impairments in memory retention and retrieval. Slow-5 is also a critical band for discriminating different mental illness states, including distinguishing between depression and mania states in patients with bipolar disorder [93], as well as differentiating schizophrenia from psychotic bipolar disorder and schizoaffective disorder [94]. Slow-4 does not show the same level of discrimination in these disorders. These findings highlight the importance of understanding slow-5 functional deficits in patients with mental disorders and identifying neuroimaging markers.

**Slow-4: Primary cognition and sensorimotor functions**

The slow-4 SSOs oscillate between approximately 0.0301 and 0.0821 Hz. Slow-4 constitutes more than half of the traditional frequency range of the rfMRI signal. For most networks, except the default mode network and the high-level visual network, slow-4 represents the main component of the power spectra compared to other frequency bands [38]. Its power spectrum density positively correlates with cognitive measures [95]. Additionally, only in slow-4, the fALFF distribution significantly is correlated with the expression of numerous genes, which are associated with various cognitive functions [96]. Although most networks are most active in the slow-4 band, their contributions to the brain's global efficiency are lower in this band compared to the lower frequency bands. Only the basal ganglia network, the visual network, and the parietal memory network have greater contributions to global efficiency in slow-4 [36]. Another study largely reproduced this finding: the global functional connections of slow-4 are sparser than those of lower frequency bands [35]. Moreover, except for the default mode network, which showed higher within-network integration in slow-5, other networks exhibited higher within-network integration in slow-4 [35]. These results suggest that, in the slow-4 band, the modularity of brain information processing is at its strongest. In general, the network organization is distinct between slow-4 and slow-5; however, in most rfMRI studies, these two bands are combined and studied as a single band. In slow-5, networks mainly play roles in integration, while their roles are more segregated in slow-4. This view is also supported by the observation that more types of brain hubs were found in slow-4 compared to other frequency bands [84]. Just as the frequency (or period) of a central processing unit (CPU) reflects computational performance, the frequency of SSOs may also reflect the performance of cognitive functions. Slow-4 exhibits the most obvious rotary pattern among the different frequency bands, implying more periodic dynamics in this band [97]. The rotary period in slow-4 is significantly slower in patients with neurodegenerative disorders, reflecting their decreased cognitive performance. In conclusion, we hypothesize that modular separation of brain functions occurs in slow-4. Specifically, the dominant functions of slow-4 are various basic cognitive functions (e.g., working memory, arithmetic, face recognition), while the function of slow-5 is dominated by more complex cognitive integration based on the functions of slow-4.

Other basic functions performed by the slow-4 SSOs include motor and sensory functions. Slow-4 SSOs primarily activate the thalamus, basal ganglia, and precentral gyrus [37,38,46]. The thalamus transmits information through numerous pathways connecting both cortical and subcortical regions to relay sensory and motor signals. All areas of the neocortex receive input from the thalamus [98,99]. The basal ganglia contain different subcortical nuclei and form rich connections with many cortical regions to perform various functions, particularly motor functions [100]. The precentral gyrus, together with the thalamus and basal ganglia, forms a highly connected neural circuit known as the cortico-basal ganglia-thalamo-cortical loop. This circuit is activated by action selection and disrupted by various motor disorders (e.g., Parkinson's disease, Huntington's disease, and ADHD) [101,102]. Considering that the network properties of the somatomotor and visual networks are integrated within and between networks at this frequency, we speculate that slow-4 SSOs are mainly involved in the long-distance transmission and integration of sensory and motor information. The sensory inputs processed at this level are integrated into different cognitive processes, supported by recent neuroimaging evidence on the cardinal symptoms of Parkinson's disease, including bradykinesia, tremor, and rigidity. Patients with this clinical condition exhibit greater alterations in SSO amplitudes within the basal ganglia, cerebellum, thalamus, and motor cortex in slow-4 compared to other frequency bands [103-106]. Furthermore, the slow-4 amplitude of SSOs in the middle frontal gyrus and inferior temporal gyrus provided the highest accuracy in discriminating patients with gait freezing from those without gait disturbances [107], as well as from healthy controls [108]. These observations of motor symptom-related clinical conditions indicate the functional associations of slow-4 SSOs.

Similar to slow-5, clinical studies have reported various abnormal SSOs in slow-4 in patients with different mental and neurological disorders. Patients with major depressive disorder present abnormal local connectivity in slow-4 and long-range connectivity in slow-5. Specifically, the slow-4 functional homogeneity metrics increased in the middle occipital gyrus and decreased in the anterior cingulate cortex, inferior/superior frontal gyrus, and bilateral thalamus [109]. No networks with significant disconnections were detected in slow-4 in treatment-sensitive patients, while in treatment-resistant patients, only the olfactory network was disconnected [67]. According to the role of this network in experiencing and processing emotion, the findings indicate that the mood-regulating mechanism is active in slow-4 and disrupted in treatment-resistant patients. Altered SSOs in the slow-4 band have been reported in patients with various diseases. Thus, it is reasonable to speculate that alterations in slow-4 SSOs are associated with various mental and neurological disorders, reflecting altered functional modularity.

**Slow-3: Vocal processing and sensory binding**

The slow-3 SSOs oscillate between 0.0821 Hz and 0.2231 Hz. Slow-3 is an important frequency band for processing vocal stimuli. Frühholz and colleagues identified several major independent components of fMRI signals related to vocal and nonvocal signal processing [110]. The third component was primarily associated with vocal signals, left-lateralized, and covered the slow-3 band. Additionally, from a temporal perspective, it appeared later than the first and second components, which covered the slow-1 and slow-2 bands, respectively, but had no preference for vocal or nonvocal signals after stimulus presentation. Thus, the third component may play a role in high-level auditory processing, especially of vocal signals. We noted that two other components identified in higher frequency bands were associated with vocal stimuli, which will be discussed later in terms of the hierarchically organized processing of vocal signals.

Sensory binding may be another functional role of slow-3 SSOs. These oscillations provided the best ability to distinguish eyes-open rfMRI scans from eyes-closed rfMRI scans [111]. In the eyes-closed condition, the sensorimotor cortex exhibited greater activation across multiple frequency bands but with broader areas of activation in slow-3 extending to the association cortex. In the absence of visual inputs, other sensory channels were more sensitive, and multimodal sensations were binding in the association cortex in slow-3. In addition, the slow-3 SSOs were the main signals that contributed to classifying schizophrenia patients from healthy controls [112]. One major symptom of schizophrenia is hallucinations, and slow-3 SSOs may participate in the manifestation of this symptom through false sensory binding. In slow-3, the cingulo-opercular network, which is linked to visual processing speed, is divided into two areas, the insula and the anterior cingulate cortex, suggesting that these two regions are functionally heterogeneous [113]. The insula exhibited the highest connectivity degree in slow-3 along with the cuneus, angular cortex, precuneus, middle cingulate cortex and inferior parietal lobe [114] while showing the highest amplitude along with the temporal cortex and subcortical regions [37, 38, 115]. The function of the insula is highly diverse, including multimodal sensory processing and sensory binding [116,117], auditory perception, emotions, and self-awareness. Overall, we speculate that the insula is involved in slow-3 and associated with the sensory binding function according to the observation that a strong positive connection exists between the insula and the visual network [113]. To date, slow-3 studies have not been sufficient to deduce its accurate physiological mechanism, and thus future studies with elaborate experimental designs are needed to test the proposed hypothesis.

**Slow-1/2: Salience detection and local integration**

The slow-2 SSOs oscillate between approximately 0.2231 Hz and 0.6065 Hz, while the slow-1 SSOs oscillate between approximately 0.6065 Hz and 1.6487 Hz. Notably, these two slow bands are rarely investigated due to the limitations of the sampling rate with fMRI and the potential overlap with physiological noise. Using fast fMRI with a sampling frequency of 1.5 Hz, a study revealed that the salience network exhibited higher power in both slow-2 and slow-1 (partial) than other networks [38]. The salience network, which largely overlaps with the ventral attention network, is presumed to be important for detecting the salience of both internal and external stimuli. It facilitates switching among various large-scale brain networks during self-monitoring and task processing [118]. Since different brain networks have been identified within different frequencies with distinct spatial extents, we postulate that the functional segregation of these networks occurs in the frequency domain. We infer that the function of the salience network in these higher slow bands is associated with external salience detection and switching to task processing. Previous EEG studies have revealed that local sensory integration involves high frequencies [6]. In the slow-2 and slow-1 bands, the superior temporal gyrus, where the auditory center resides, and the visual cortex are more active than other regions [47]. The activity in these primary sensory areas may signify that the initial processing of sensory stimuli occurs at high frequencies. In the auditory experiment discussed in the previous section, researchers identified two sets of functional components in the slow-1 and slow-2 bands [110]. The first set of components represents sustained BOLD patterns occurring immediately after the presentation of stimuli, covering the slow-1 (partial) and slow-2 bands, respectively. The second set of components is related to transient oscillations that appear sequentially after the stimuli and are located in the primary, secondary, and higher-level auditory cortex. Therefore, these components may represent information flow through the auditory cortex. Additionally, a sustained component associated only with vocal stimuli has been identified in slow-2. It appears much earlier than the component identified in slow-3, which is preferentially activated by vocal stimuli. Thus, it may represent the initial processing of vocal signals, while the slow-3 component may be involved in higher-level processing. In summary, the primary processing of both vocal and nonvocal signals occurs in the slow-1/2/3 bands and is hierarchically organized in the frequency domain. The slow-1 and slow-2 bands mainly perform initial processing and local integration, while the exact activated frequencies depend on the frequency of stimuli. In contrast, the slow-3 band is involved in semantic processing and sensory binding. We further infer that the primary processing of visual signals also occurs in these higher-frequency slow bands with similarly organized hierarchies.

## A Three-level SSO Integrative Model

Based on the abovementioned empirical findings, we propose a novel three-level integrative model of N3L-guided SSOs (Figure 5) to elucidate their functions and interrelationships. This model categorizes the six (slow-6 to slow-1) BOLD SSO bands into three fundamental levels of functional stream into constructing the complex tapestry or waves of neural activity.

The first level is detection (slow-1/2/3), where SSOs are faster and primarily driven by streams to external inputs, and unimodal brain areas are most active. According to the specific functions of the different bands, the detection level is further divided into two components responsible for sensation and perception. Specifically, slow-1 and slow-2 oscillations belong to the sensation component, responsible for detecting external signals and receiving and initially processing sensory information locally. The exact activation frequency during perception is determined by the physical characteristics of external signals. Since the SSOs of these two bands fluctuate quickly, they can adapt to the rapidly changing environments. Slow-3 SSOs are assigned to the perception component, associated with sensory binding and perception. Semantic processing may occur within this frequency range. Generally, SSOs at the detection level constitute a subsystem for the preliminary processing of external signals. These functions are essential for an individual to survive while responding to constantly changing environments.

The second level is computation or representation (slow-4), where SSOs control sensorimotor functions and multiple basic cognitive functions through computation across different network modules with various spatial distribution profiles and connection patterns. SSOs show the strongest modular activity pattern at this level. Sensory information from the detection level is further integrated and transferred among the modules at this level to participate in various cognitive processes. The computation level functions like a central processing unit in a computer: it receives inputs from higher-frequency slow bands and is modulated by lower-frequency slow bands to implement various calculations on demand for informatic representations. This characteristic implies that it may function as the core of human intelligence.

The third level is modulation (slow-5 and slow-6), where SSOs are internally driven and perform top-down functions distributed across a wide range of brain areas. Attention control and fast modulation are one of the main functions of SSOs at this level. Specifically, slow-5 SSOs are related to consciousness, memory, and complex cognitive functions, integrating information flow from the computation level and activating the default mode network as its core network. Meanwhile, slow-6 SSOs participate in broader integration of information. SSOs of this level regulate attention cycles through slow waves. Ultraslow oscillations achieve the system's adaptive regulation by modulating higher-frequency oscillations. Briefly, the modulation level prepares an individual to stay conscious and self-aware, enabling them to deal with the complexity of responding to the environment as a predictive brainy machine.

Generally, BOLD SSOs in different slow bands dominate distinct roles in formatting the human brain's intrinsic function and interact to maintain a hierarchical dynamic system as its neuronal orchestra. This system simultaneously accommodates both the scale-specific and scale-free characteristics of neural activity. BOLD oscillations in different frequency bands serve distinct functions while interacting and providing feedback across different hierarchical levels, demonstrating the multi-scale behavior of a dynamic system. High-frequency SSOs are more functionally segregated and activated more locally (through short-distance connections). Low-frequency SSOs are more functionally integrated and activated more remotely (through long-distance connections and synchronization) [119]. The organization of brain oscillations in this manner has physiological implications and neurological underpinnings. Thin and less-myelinated nerve fibers connect multimodal areas, while thick and myelinated nerve fibers connect unimodal areas [7]. The thicker and more myelinated the nerve fibers are, the faster the conduction speed of information flow. This finding may partially explain why high-order functions dominate low-frequency SSOs, while primary functions dominate high-frequency SSOs. An interesting analogy is to compare this model to a computer. The detection level corresponds to the input device of the computer, which detects external signals and transfers them into neural signals. The computation or representation level corresponds to the central processing and memory units of the computer. These units receive the neural signals from the detection level and extract long-term memory from the modulation level to perform various operations based on instructions. The modulation level is like the hard drive and operating system that store long-term memory and regulate brain signals at the other two levels.

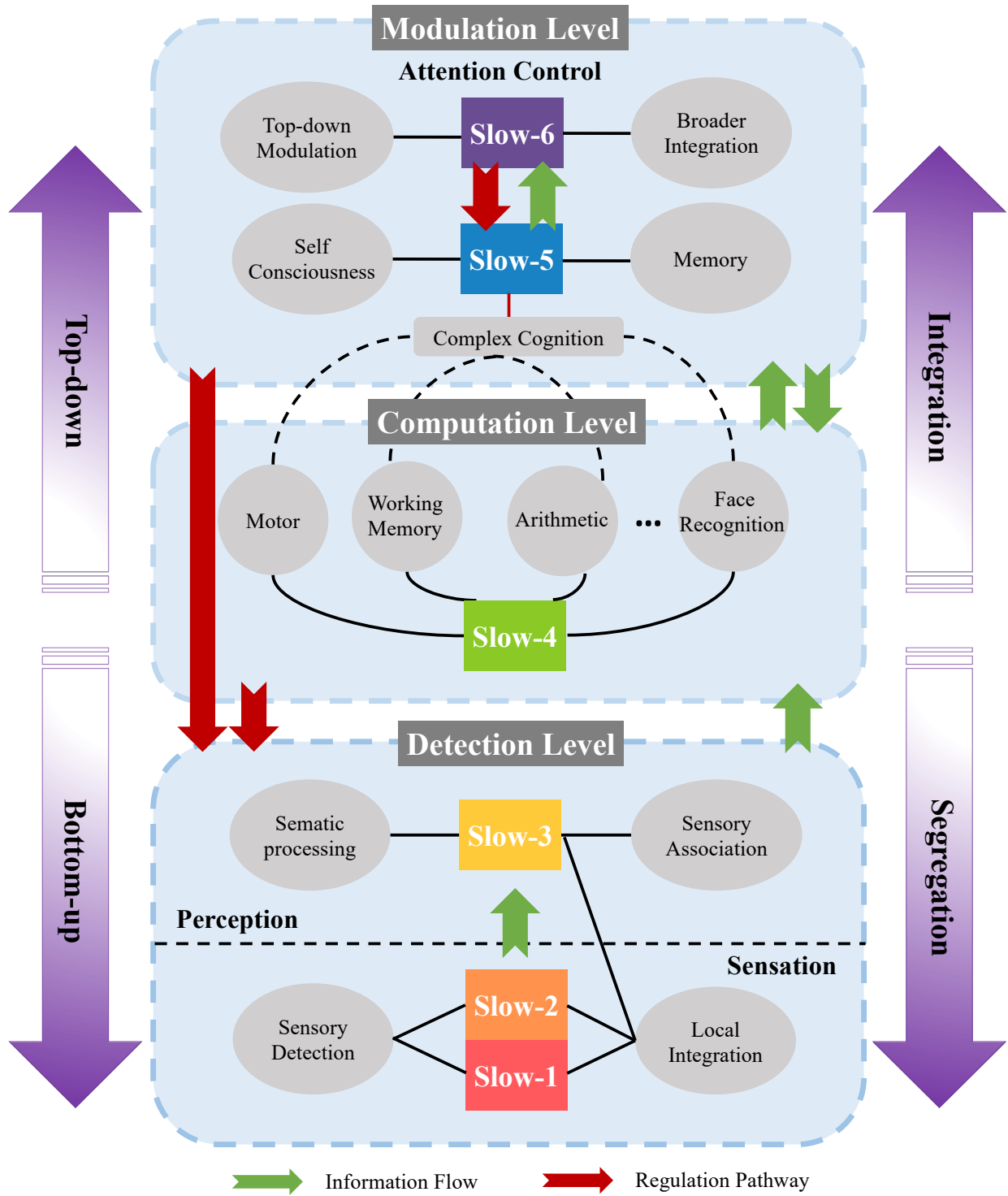

*Figure 5. A three-level SSO integrative model*. The model categorizes the six slow bands of SSOs into three interactive levels: detection, computation and modulation. The detection level (slow-1/2/3) is bottom-up driven and responsible for sensation and perception. The computation or representation level (slow-4) is the core intelligence with multiple functional modules. The modulation level (slow-5/6) integrates information from higher frequencies and modulates faster SSOs.

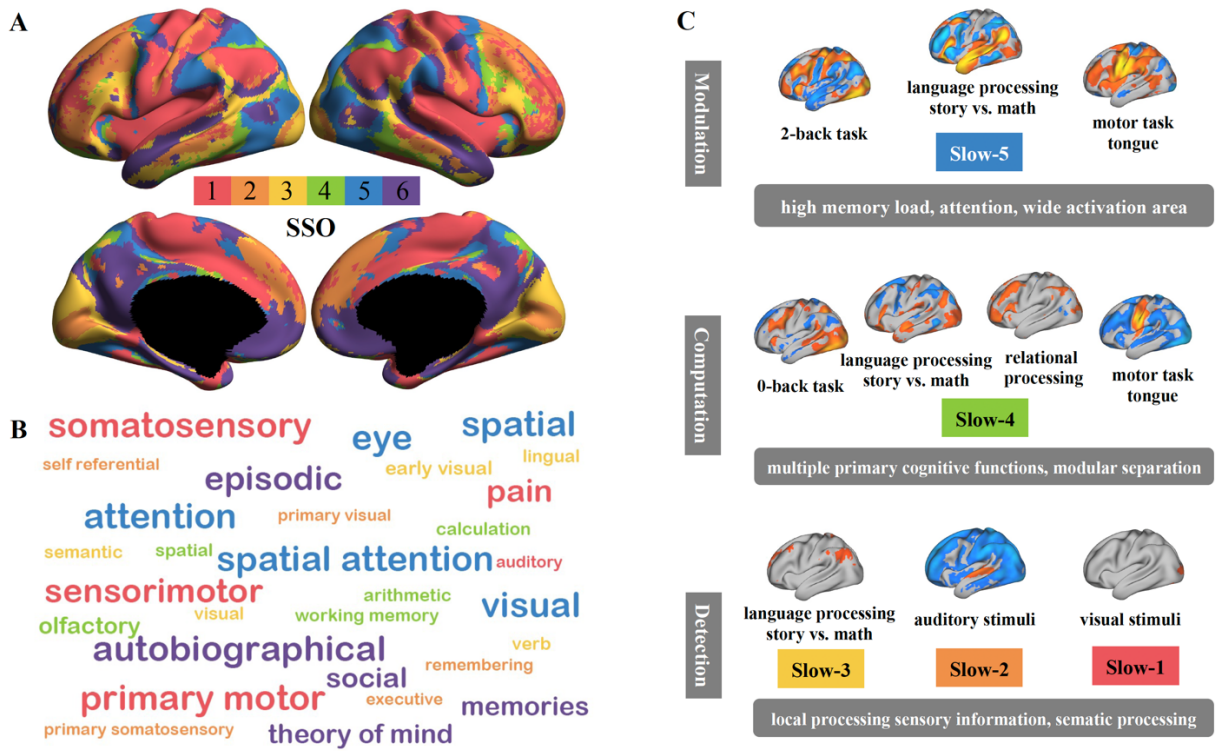

*Figure 6. Evidence supporting the three-level SSO integrative model from both spontaneous and task-induced brain activity*. *(A) The frequency-rank map of the highest values for the first gradient reflects the frequency band in which a particular brain area has the highest level of integration. (B) Meta-analytic decoding of the frequency-rank map. Font color represents the frequency band associated with the functional term, and font size reflects the correlation coefficient size. (C) Multi-band activation patterns of different tasks align with the three-level SSO integrative model. At the detection level, activation is limited to the sensory regions, especially in slow-1 and slow-2. At the computation or representation level, different tasks elicit unique sets of activation regions. At the modulation level, high-memory load tasks have stronger activation effects. Attention networks are widely activated across tasks.*

This model is based on empirical findings from both basic and clinical MBFA studies. It reveals the unique frequency characteristics of BOLD SSOs and is consistent with neural oscillations measured using other techniques. We argue that this model is derived from both theoretical and empirical evidence and could help fill the gap between observed data and underlying mechanisms of brain rhythms, as well as transformations from basic findings to clinical applications. More testable experiments are needed to further enrich, modify, and improve this initial slow-band oscillation model. To verify the validity of this model, we conducted a preliminary verification of the function of

BOLD oscillations in six frequency bands from the perspectives of spontaneous brain activity and task-induced brain activity.

**Primary evidence from spontaneous brain activity**

The cortex's processing hierarchy is embedded in spontaneous neural activity. Using dimension reduction methods on the functional connection matrix, researchers have found a set of FC gradient components, among which the first gradient depicts the hierarchical organization across the cortex [120]. Along the first gradient, brain regions are arranged in order from primary unimodal regions to high-level transmodal regions. The unimodal end is anchored in the primary sensorimotor regions, and regions of the transmodal end have the highest integration processing level. Thus, for a certain region on the cortex, its position on the first gradient axis reflects the hierarchical procession level. For instance, brain regions near the transmodal end of the axis would have higher integration processing levels than regions near the unimodal end. Based on this, in a recent study, we explored the frequency functional hierarchy of SSOs by mapping the frequency of the highest integration level across the cortex [121]. We first computed the FC gradients for six frequency bands (slow-1 to 6) and normalized the first gradient for each frequency band so that the first gradients could be comparable among the bands. Then, for each vertex, the first gradient values across the bands were ranked. The higher the gradient value of a vertex in a certain frequency band, the closer the position of the vertex in the first gradient in the frequency band is to the transmodal end. A distribution map of the frequency bands with the highest-ranking value across the cortex was then obtained (Figure 6A). This map reflected the frequency band at which brain regions exhibit the highest level of integration. The map showed significant clustered patterns that approximately outlined the resting-state networks. Primary networks exhibited the highest level of integration in high-frequency bands, while high-level networks showed divergent patterns. A meta-analytic decoding of those clustered regions in each frequency band was further conducted (Figure 6B). In summary, high-frequency-band regions, namely, regions showing the highest integration level at the detection level, are mostly associated with sensory, motor and language functions, except those in slow-2. The computation level (slow-4) regions are mostly linked to executive functions. The low-frequency-band regions, that is, regions exhibiting the highest integration level at the modulation level, are mostly related to bottom-up and self-related functions. These results are basically consistent with the hypothesis of the function of each frequency band in the three-level integration model. This suggests that in the seemingly chaotic spontaneous brain activity, the functional integration level of brain oscillations exists not only at the spatial level but also at the time or frequency level. Therefore, combining spatial and frequency information to decode the function of the cerebral cortex and brain oscillations provides a more comprehensive view of the brain.

**Primary evidence from task-induced brain activity**

Spontaneous brain activity accounts for approximately 70% of the brain's energy consumption, while task-evoked brain activation, which induces blood flow changes, accounts for only 5% or less [122,123]. Therefore, the resting state of the brain is not truly at rest; instead, it harbors 'dark energy' rich with information. The brain's spontaneous neural activity encompasses patterns linked to various cognitive processes, and individual differences in cognitive task activation patterns can be predicted through resting-state BOLD signals [124]. As a result, the present three-level SSO integrative model is not only applicable to but can also be validated by task-evoked activity. Task activation analysis constitutes the most direct way to locate brain activity induced by task manipulation. Traditional task activation studies primarily focus on the spatial location of induced brain activity. By conducting task activation analysis across multiple frequency bands, we can identify both frequency and spatial activations related to task manipulation. In a previous study, we mapped several cognitive processes across frequency bands using multi-band activation analysis on seven task-fMRI scans from the HCP dataset [125]. The multi-band activation patterns align with the assumptions of the three-level integrative model (Figure 6C). Firstly, we posited that cognitive processes are supported by multiple frequency bands of brain activity; therefore, we assumed that each task would elicit activations in more than one band. The results confirmed this hypothesis, demonstrating that every task significantly activated brain regions in multiple frequency bands. Secondly, we assumed that BOLD oscillations at the detection level are mainly involved in the primary processing of sensory information. The results showed that in the slow-1 and slow-2 bands, all tasks activated brain regions primarily within the primary sensory cortex. Specifically, tasks using visual stimuli activated the visual cortex, while tasks using auditory stimuli activated the auditory cortex. In the slow-3 band, in addition to the sensory cortex, some tasks also induced activations in small regions of the attention networks. These observations indicate that at the detection level, brain activity induced by task manipulation is mainly related to local processing of sensory information, and that in the slow-3 band, the primary integration of sensory information occurs. Thirdly, the spatial activation patterns of different tasks diverged at the computation level as expected. Different conditions within the motor task caused activation of the motor cortex corresponding to each body part. Each of the other cognitive tasks individually activated a set of brain regions associated with the task. The tasks exhibited specific and modular activation patterns in the slow-4 band, indicating that various basic cognitive processes are carried out by activating their own unique task networks at the computation level. Lastly, we hypothesized that the modulation level is associated with high cognitive load processing and advanced integration level processing. In general, the results showed that the same task condition had a wider activation range in the slow-5 band, and a large area of the attention networks was activated in this band. This suggests that cognitive processing in the slow-5 band involves higher levels of integration and regulation. Particularly, in the working memory task, the 2-back vs. 0-back contrast only showed significant activation in the slow-5 band, and the precuneus, a region related to memory functions, was only activated in the slow-5 band during the 2-back condition and 2-back vs. 0-back contrast. This indicates that high memory load processing occurs in the slow-5 band. Unfortunately, due to the short scanning time of each task, we were unable to verify the activation mode of the slow-6 band during the task state. Still, through the multi-band activation analysis of task-induced brain activity, we essentially verified the hypothesis of the three-level integrative model regarding the functions of BOLD oscillations at each level. In future studies, the use of task paradigms specifically designed for multi-band studies may more accurately reveal the function of BOLD oscillations in each frequency band and their mechanisms of activity in cognitive processing.

# Discussion

Low-frequency oscillations of slow spontaneous activity are organized into a high-order complex system in the human brain. This organization represents an intrinsic functional hierarchy that is congruent with the anatomical hierarchy in both spatial and temporal dimensions and has been described in an increasing number of studies published in recent years [126]. Our review provides additional insights into the intrinsic architecture of the complex human brain system from a hierarchical model of its spontaneous frequencies.

## SSOs organize brain function into a frequency hierarchy

Leveraging gradient mapping approaches, Daniel Margulies and colleagues revealed a set of connectivity gradients from the functional connectome of the brain on a macroscale of space [120]. Specifically, a gradient anchored by high-level transmodal associative areas at one end and unimodal primary areas at the other end explained the highest variance in the observed connectivity matrix and thus was named the principal gradient. This functional gradient is organized in accordance with the anatomical hierarchy, encoding the microstructure gradient changes from the sensorimotor unimodal areas toward default transmodal areas [127]. From the perspective of information processing on the temporal scale, the temporal span of information accumulation in the brain shows a gradient structure constrained by the anatomy that is consistent with the functional hierarchy. Specifically, the length of "the temporal receptive windows" increases from primary sensorimotor areas (simple, quick processing of unimodal information by neuronal connections with more myelination) to high-order associative areas (complex slow processing of transmodal information by neuronal connections with less myelination) [128]. Overall, on both the temporal and spatial scales, the human brain organizes SSOs to achieve the functional integration of primary and associative units in information processing according to the gradients of the functional and structural connectomes.

We proposed a hierarchical model of brain SSOs across frequency scales incorporating both spatial and temporal information, reflecting the underlying anatomical basis and information flow processes throughout the brain. The organization of SSOs exposes the spatial hierarchy of a particular frequency band at every interval of the frequency scale. In other words, at high-frequency scales, the SSOs are mainly distributed in the primary areas where connections are more local. In contrast, at low-frequency scales, the SSOs are mainly assembled in associative areas where connections are more distant. Notably, most studies examining the traditional frequency band (0.01-0.1 Hz) are considered a special case of MBFA. In the frequency domain, information about the time dimension is encoded at various frequencies. A short "temporal receptive window" is related to the characteristics of a high-frequency oscillation for transferring information quickly, which predominates in the sensorimotor areas. In contrast, a low-frequency oscillation, which predominates in the associative areas, provides a long "temporal receptive window" to analyze and integrate information slowly. This frequency organization receives good support from the structural hierarchy, in which the diameter and myelination of the fibers decrease from the primary sensorimotor regions to high-level associative regions [7]. From the perspective of frequency, the information flow throughout the brain becomes more pronounced. In the three-level integrative model, faster oscillations generated in unimodal primary sensory regions allow external information to enter the detection level to be processed and combined with internal knowledge at the computation level and modulation level by slower oscillations generated in transmodal high-order associative regions. This information, in addition to conventional single-band research, provides an innovative angle to pinpoint the pathway of information flow through both the spatial and temporal dimensions. Decoding the perception of space and time is a key long-term topic in cognitive neuroscience, similar to the notions of space and time, the two crucial dimensions in physics [129]. Frequency-dependent studies of brain mechanisms may thus be one of the turning points in future research to understand how the brain accomplishes spatiotemporal perception across different frequencies.

**Lifespan development of SSOs through a frequency lens**

Characterizing lifespan development is crucial for research into both healthy brain mechanisms and clinical etiology [130-132]. Divergent trajectories of lifespan development have been presented for various brain structures, functions, and psychological traits [133], but rarely for measurements across multiple frequency bands. We argue that oscillations at different frequency bands predominate in different functions; therefore, the functional organization patterns of brain oscillations in various frequency bands might also develop differently over the lifespan. Compared to the other two levels in the proposed model, the detection level is linked to more fundamental abilities, including sensation, perception, and language comprehension. Thus, we speculate that functional maturation at this level starts earlier than at the other levels. The computation level is the foundation of human brain intelligence and is connected to motor and diverse cognitive functions. Understanding how brain oscillations at this level evolve has major implications for understanding how children's intellectual development occurs in the brain. For ageing, which is vulnerable to cognitive decline, particularly working memory deficits [134,135], the related oscillations at this level may be the first to degenerate functionally in an ageing brain. A study revealed that global homogeneity and homotopy of SSOs at the computation level increased with age during normal ageing [136]. This may reflect a decrease in the modularity of the computation level SSOs during ageing, requiring more brain regions to process a task together. The modulation level is associated with long-term memory and top-down functions, such as attention, consciousness and self-consciousness. Understanding of the working mechanisms underlying the development of these oscillations in the brain may be obtained by examining how intrinsic activity and connection properties develop at modulation frequencies. Overall, research on the multifrequency band trajectories of brain development across the lifespan is quite sparse. The benefits of association studies between mental or behavioral traits and low-frequency oscillations in various frequency bands, aiming to reveal critical stages of development and ageing, remain elusive.

Mental disorders may be conceptually assessed and quantified as deviations from normal brain development at various stages of the lifespan. Numerous studies have documented the frequency dependence of SSO changes linked to mental disorders. More than half of all mental disorders develop before adulthood [137]. The trajectories of healthy brain development from different frequency bands provide normative references for research on the mechanisms underlying brain development during the atypical conditions associated with neurological and psychiatric disorders. On the one hand, a more effective approach might be to identify biomarkers by concentrating on some particular frequency-bands that are highly sensitive to dimensional symptoms. For instance, ADHD patients exhibit more alterations in slow-6 oscillations than other bands [61]. Therefore, an examination of changes connected to the disorder in the slow-6 frequency range rather than the traditional frequency range would be better to identify biomarkers for ADHD. On the other hand, the use of all frequency bands rather than only a single frequency range is more informative and comprehensive for disease classification and screening. More information will be obtained by decomposing a broadband signal into oscillations produced by several physiological mechanisms. In the composite frequency band, information from a single frequency band, particularly phase-related information, may be concealed or cancelled. As in the famous Simpson's paradox, compound signals sometimes show a pattern that is opposite to that of the subbands. Thus, the multifrequency band approach is more informative and more obvious in machine learning. The classification accuracy increases with the number of frequency bands used to classify various fMRI task conditions. The classification outcome that includes every frequency band performs much better than the original signal without frequency decomposition [138]. Similarly, in clinical situations, the multi-band frequency approach outperforms the single-band frequency approach in terms of classification performance when distinguishing schizophrenia patients from healthy controls [112]. In summary, using the development trajectories of multifrequency oscillations as normative references, the multifrequency band approach potentially provides more resources for timely and accurate diagnosis, early warning and intervention in mental disorders [139].

**Methodological and technical considerations**

Several opportunities, limitations and challenges of the multifrequency band approaches proposed must be considered. The performance of scanners has significant effects on the frequency bands of low-frequency oscillations of interest. A portion of the slow-1 band is currently covered at the highest frequency by the BOLD signal produced using multiband fast scanning technology. This approach provides a fantastic opportunity to examine the functions of various oscillations by leveraging the multifrequency band method. In the future, as scanning technology advances, we will be able to identify and investigate BOLD signals at much higher frequencies [140]. At present, the multifrequency investigation of BOLD-derived oscillations is still in its infancy. Notably, most of the findings we reviewed were obtained from clinical and rfMRI studies. More concrete data are needed, particularly those from multifrequency band analyses of task-based or naturalistic fMRI, to confirm and further pinpoint the precise roles of these various frequency bands in the hierarchical model. Accordingly, a better experimental paradigm is needed for multifrequency band methodology. For both resting and task fMRI, as well as the recently proposed naturalistic fMRI [141], longer sampling times and smaller sampling intervals are recommended to cover a wider frequency range. The highest frequency bands provided by the Human Connectome Project fMRI protocol [30] cover a small portion of slow-1 with a sample rate of 720 ms, which is the typical frequency for multiband fMRI scanning, and reaches approximately the highest frequency at 0.6944 Hz. Meanwhile, a 9-

minute scan can reliably reach the lower limit of slow-5, while a 24-minute sampling duration can reliably cover the full range of slow-6. Accordingly, we advise studying the entire frequency range, if possible, to acquire a comprehensive understanding of how the brain functions based on a comprehensive investigation of slow oscillations. However, the cost increases with the length of the scan. Therefore, researchers occasionally must choose between the scanning time and the lower limit of the frequency. In that case, the three-level integrative model we proposed can be used to select bands to investigate specific functions or relevant brain disorders. Finally, as mentioned earlier, electrophysiological signals and BOLD signals are measurements of neural activity at different scales. Therefore, we believe there may be a correspondence between these two types of signals across the large-scale frequency hierarchy, which reflects the self-similarity of neural activity as a complex system [9]. The potential correspondence between BOLD signals and electrophysiological activity across different frequency bands remains an open question. The three-level SSO integrative model provides a reference for unlocking this correspondence. Future research on this topic will contribute to a deeper, multi-scale understanding of the mechanisms underlying neural activity.

## Conclusions

We see that the brain oscillates slowly and spontaneously as a high-order, elaborated system by leveraging the BOLD-rfMRI. Over the past decade, gradual advances in our understanding of this hierarchical system based on slow-frequency bands with distinct oscillation patterns and associated clinical conditions have been achieved. This review summarizes and disassembles the evidence from previous studies characterizing the functions of each band. A three-level integrative model emerged from these observations, illuminating how these slow bands are segregated and integrated in complex brain systems to achieve different levels of brain oscillations in the organizational hierarchy. With potential insights of this model into the mechanisms on brain's dark energy, this review provides a theoretical framework to guide the interpretation of findings and design of experiments for both basic and clinical research, representing a paradigm shift for future human brain mapping studies [142-145].

## Acknowledgements

This work has been supported by the scientific and technological innovation 2030 - the major project of the Brain Science and Brain-Inspired Intelligence Technology (2021ZD0200500) and the Interdisciplinary Brain Database for In-vivo Population Imaging (ID-BRAIN) at the National Basic Science Data Center. We thank the Research Program on Discipline Direction Prediction and Technology Roadmap of China Association for Science and Technology for bibliometric resources, and the National Basic Science Data Center for informatics resources.

## Conflict of interest

The authors declare that they have no conflict of interest.

## Availability of data and materials

The single-subject rfMRI data is available from the database of the Human Connectome Project (HCP) (https://db.humanconnectome.org). which is supported by the NIH Blueprint for Neuroscience Research 1U54MH091657 (principal investigators: David Van Essen and Kamil Ugurbil) and the McDonnell Center for Systems Neuroscience at Washington University.

## Code availability

All codes we used for generating figures are shared via the GitHub repository of the Connectome Computation System at https://github.com/zuoxinian/CCS.

## Authors' contributions

X.N.Z. designed the study. Z.Q.G. and X.N.Z. conducted analyses, wrote and edited the manuscript as well are involved in all other aspects of contribution.

## References

[1] Bullock TH. Signals and signs in the nervous system: The dynamic anatomy of electrical activity is probably information-rich. Proc Natl Acad Sci U S A 1997;94:1–6.

[2] Buzsaki G, Voroslakos M. Brain rhythms have come of age. Neuron 2023;111:922–926.

[3] Berger H. Electroencephalogram in humans. Arch Psychiatr Nervenkr 1929;87:527–570.

[4] Klimesch W. EEG alpha and theta oscillations reflect cognitive and memory performance: A review and analysis. Brain Res Rev 1999;29:169–195.

[5] Engel AK, Fries P, Singer W. Dynamic predictions: Oscillations and synchrony in top-down processing. Nat Rev Neurosci 2001;2:704–716.

[6] von Stein A, Sarnthein J. Different frequencies for different scales of cortical integration: From local gamma to long range alpha/theta synchronization. Int J Psychophysiol 2000;38:301–313.

[7] Aboitiz F. Brain connections: Interhemispheric fiber systems and anatomical brain asymmetries in humans. Biol Res 1992;25:51–61.

[8] Glasser MF, Goyal MS, Preuss TM, Raichle ME, Van Essen DC. Trends and properties of human cerebral cortex: Correlations with cortical myelin content. Neuroimage 2014;93:165–175.

[9] Munn BR, Muller EJ, Favre-Bulle I, Scott E, Lizier JT, Breakspear M, et al. Multiscale organization of neuronal activity unifies scale-dependent theories of brain function. Cell 2024.

[10] Gallos LK, Song C, Makse HA. A review of fractality and self-similarity in complex networks. Physica A 2007;386:686–91.

[11] Anderson PW. More is different. Science 1972;177:393–396.

[12] Raichle ME. The brain's dark energy. Science 2006;314:1249–1250.

[13] Arthurs OJ, Boniface S. How well do we understand the neural origins of the fMRI BOLD signal? Trends Neurosci 2002;25:27–31.

[14] Rees G, Howseman A, Josephs O, Frith CD, Friston KJ, Frackowiak RSJ, et al. Characterizing the relationship between BOLD contrast and regional cerebral blood flow measurements by varying the stimulus presentation rate. Neuroimage 1997;6:270–278.

[15] Kim T, Hendrich KS, Masamoto K, Kim SG. Arterial versus total blood volume changes during neural activity-induced cerebral blood flow change: Implication for BOLD fMRI. J Cereb Blood Flow Metab 2007;27:1235–1247.

[16] Kim SG, Ogawa S. Biophysical and physiological origins of blood oxygenation level-dependent fMRI signals. J Cereb Blood Flow Metab 2012;32:1188–1206.

[17] Arthurs OJ, Williams EJ, Carpenter TA, Pickard JD, Boniface SJ. Linear coupling between functional magnetic resonance imaging and evoked potential amplitude in human somatosensory cortex. Neuroscience 2000;101:803–806.

[18] Heeger DJ, Huk AC, Geisler WS, Albrecht DG. Spikes versus BOLD: What does neuroimaging tell us about neuronal activity? Nat Neurosci 2000;3:631–633.

[19] Rees G, Friston KJ, Koch C. A direct quantitative relationship between the functional properties of human and macaque V5. Nat Neurosci 2000;3:716–723.

[20] Logothetis NK, Pauls J, Augath M, Trinath T, Oeltermann A. Neurophysiological investigation of the basis of the fMRI signal. Nature 2001;412:150–157.

[21] Biswal B, Yetkin FZ, Haughton VM, Hyde JS. Functional connectivity in the motor cortex of resting human brain using echo-planar MRI. Magn Reson Med 1995;34:537–541.

[22] Biswal BB. Resting state fMRI: A personal history. Neuroimage 2012;62:938–944.

[23] Fox MD, Raichle ME. Spontaneous fluctuations in brain activity observed with functional magnetic resonance imaging. Nat Rev Neurosci 2007;8:700–711.

[24] Biswal B, DeYoe EA, Hyde JS. Reduction of physiological fluctuations in fMRI using digital filters. Magn Reson Med 1996;35:107–113.

[25] Lowe MJ, Mock BJ, Sorenson JA. Functional connectivity in single and multislice echoplanar imaging using resting-state fluctuations. Neuroimage 1998;7:119–132.

[26] Power JD, Schlaggar BL, Petersen SE. Studying brain organization via spontaneous fMRI signal. Neuron 2014;84:681–696.

[27] Cohen AL, Fair DA, Dosenbach NUF, Miezin FM, Dierker D, Van Essen DC, et al. Defining functional areas in individual human brains using resting functional connectivity MRI. Neuroimage 2008;41:45–57.

[28] Achard S, Bullmore ET. Efficiency and cost of economical brain functional networks. PLoS Comput Biol 2007;3:174–183.

[29] Van Essen DC, Smith SM, Barch DM, Behrens TEJ, Yacoub E, Ugurbil K, et al. The Wu-Minn human connectome project: An overview. Neuroimage 2013;80:62–79.

[30] Smith SM, Beckmann CF, Andersson J, Auerbach EJ, Bijsterbosch J, Douaud G, et al. Resting-state fMRI in the human connectome project. Neuroimage 2013;80:144–168.

[31] Yan L, Zhuo Y, Ye Y, Xie SX, An J, Aguirre GK, et al. Physiological origin of low-frequency drift in blood oxygen level dependent (BOLD) functional magnetic resonance imaging (fMRI). Magn Reson Med 2009;61:819–827.

[32] Boubela R, Kalcher K, Huf W, Kronnerwetter C, Filzmoser P, Moser E. Beyond noise: Using temporal ICA to extract meaningful information from high-frequency fMRI signal fluctuations during rest. Front Hum Neurosci 2013;7:168.

[33] Hu R, Peng Z, Zhu X, Gan J, Zhu Y, Ma J, et al. Multi-band brain network analysis for functional neuroimaging biomarker identification. IEEE Trans Med Imaging 2021;40:3843–3855.

[34] Yu Q, Cai Z, Li C, Xiong Y, Yang Y, He S, et al. A novel spectrum contrast mapping method for functional magnetic resonance imaging data analysis. Front Hum Neurosci 2021;15.

[35] Thompson WH, Fransson P. The frequency dimension of fMRI dynamic connectivity: Network connectivity, functional hubs and integration in the resting brain. Neuroimage 2015;121:227–242.

[36] Park YH, Cha J, Bourakova V, Lee JM. Frequency specific contribution of intrinsic connectivity networks to the integration in brain networks. Sci Rep 2019;9:4072.

[37] Baria AT, Baliki MN, Parrish T, Apkarian AV. Anatomical and functional assemblies of brain BOLD oscillations. J Neurosci 2011;31:7910–7919.

[38] Gohel SR, Biswal BB. Functional integration between brain regions at rest occurs in multiple-frequency bands. Brain Connect 2015;5:23–34.

[39] Mierau A, Klimesch W, Lefebvre J. State-dependent alpha peak frequency shifts: Experimental evidence, potential mechanisms and functional implications. Neuroscience 2017;360:146–154.

[40] Palva JM, Palva S. Functional integration across oscillation frequencies by cross-frequency phase synchronization. Eur J Neurosci 2018;48:2399–2406.

[41] Canolty RT, Ganguly K, Kennerley SW, Cadieu CF, Koepsell K, Wallis JD, et al. Oscillatory phase coupling coordinates anatomically dispersed functional cell assemblies. Proc Natl Acad Sci U S A 2010;107:17356–17361.

[42] Mather M, Nga L. Age differences in thalamic low-frequency fluctuations. Neuroreport 2013;24:349–353.

[43] Penttonen M, Buzsáki G. Natural logarithmic relationship between brain oscillators. Thalamus Relat Syst 2003;2:145–152.

[44] Di Martino A, Ghaffari M, Curchack J, Reiss P, Hyde C, Vannucci M, et al. Decomposing intra-subject variability in children with attention-deficit/hyperactivity disorder. Biol Psychiatry 2008;64:607–614.

[45] Klimesch W. The frequency architecture of brain and brain-body oscillations: An analysis. Eur J Neurosci 2018;48:2431–2453.

[46] Zuo XN, Di Martino A, Kelly C, Shehzad ZE, Gee DG, Klein DF, et al. The oscillating brain: Complex and reliable. Neuroimage 2010;49:1432–1445.

[47] Gong ZQ, Gao P, Jiang C, Xing XX, Zuo XN. DREAM: A toolbox to decode rhythms of the brain system. Neuroinformatics 2021;19:529–545.

[48] Klar P, Çatal Y, Langner R, Huang ZR, Northoff G. Scale-free dynamics in the core-periphery topography and task alignment decline from conscious to unconscious states. Commun Biol 2023;6:48.

[49] Klar P, Çatal Y, Langner R, Huang ZR, Northoff G. Scale-free dynamics of core-periphery topography. Hum Brain Mapp 2023;44:1997–2017.

[50] Buzsaki G, Draguhn A. Neuronal oscillations in cortical networks. Science 2004;304:1926–1929.

[51] Lombardi F, De Martino D. Oscillations and avalanches coexist in brain networks close to criticality. Nat Comput Sci 2023;3:194-195.

[52] Lombardi F, Pepic S, Shriki O, Tkacik G, De Martino D. Statistical modeling of adaptive neural networks explains co-existence of avalanches and oscillations in resting human brain. Nat Comput Sci 2023;3:254-263.

[53] He BJ. Scale-free brain activity: Past, present, and future. Trends Cogn Sci 2014;18:480–487.

[54] Gerster M, Waterstraat G, Litvak V, Lehnertz K, Schnitzler A, Florin E, et al. Separating neural oscillations from aperiodic 1/f activity: Challenges and recommendations. Neuroinformatics 2022;20:991-1022.

[55] Zuo X-N, Xu T, Milham MP. Harnessing reliability for neuroscience research. Nat Hum Behav 2019;3:768–771.

[56] Milham MP, Vogelstein J, Xu T. Removing the reliability bottleneck in functional magnetic resonance imaging research to achieve clinical utility. JAMA Psychiatry 2021;78:587–588.

[57] Schaefer A, Kong R, Gordon EM, Laumann TO, Zuo X-N, Holmes AJ, et al. Local-global parcellation of the human cerebral cortex from intrinsic functional connectivity MRI. Cereb Cortex 2018;28:3095–3114.

[58] Du J, DiNicola LM, Angeli PA, Saadon-Grosman N, Sun W, Kaiser S, et al. Organization of the human cerebral cortex estimated within individuals: networks, global topography, and function. J Neurophysiol 2024;131:1014–1082.

[59] Dosenbach NUF, Raichle ME, Gordon EM. The brain's action-mode network. Nat Rev Neurosci 2025;

[60] Yeo BT, Krienen FM, Sepulcre J, Sabuncu MR, Lashkari D, Hollinshead M, et al. The organization of the human cerebral cortex estimated by intrinsic functional connectivity. J Neurophysiol 2011;106:1125-1165.

[61] Kong R, Yang Q, Evan G, Xue A, Yan X, Csaba O, et al. Individual-specific areal-level parcellations improve functional connectivity prediction of behavior. Cereb Cortex 2021;31:4477–4500.

[62] Smith AM, Lewis BK, Ruttimann UE, Ye FQ, Sinnwell TM, Yang Y, et al. Investigation of low frequency drift in fMRI signal. Neuroimage 1999;9:526–533.

[63] Nunez PL. Toward a quantitative description of large-scale neocortical dynamic function and EEG. Behav Brain Sci 2000;23:371–398.

[64] Yu X, Yuan B, Cao Q, An L, Wang P, Vance A, et al. Frequency-specific abnormalities in regional homogeneity among children with attention deficit hyperactivity disorder: A resting-state fMRI study. Sci Bull 2016;61:682–692.

[65] Zhang T, Xu P, Guo L, Chen R, Zhang R, He H, et al. Multivariate empirical mode decomposition based sub-frequency bands analysis of the default mode network: A resting-state fMRI data study. Appl Informatics 2015;2:2.

[66] Yue Y, Jia X, Hou Z, Zang Y, Yuan Y. Frequency-dependent amplitude alterations of resting-state spontaneous fluctuations in late-onset depression. Biomed Res Int 2015;2015:505479.

[67] He Z, Cui Q, Zheng J, Duan X, Pang Y, Gao Q, et al. Frequency-specific alterations in functional connectivity in treatment-resistant and -sensitive major depressive disorder. J Psychiatr Res 2016;82:30–39.

[68] Pang Y, Cui Q, Wang Y, Chen Y, Yang Q, Chen H. Major depressive disorder shows frequency-specific abnormal functional connectivity patterns associated with anhedonia. In: Proceedings of the Third International Conference on Medical and Health Informatics 2019. Xiamen, China: Association for Computing Machinery; 2019. p. 65–70.

[69] Martino M, Magioncalda P, Yu H, Li X, Wang Q, Meng Y, et al. Abnormal resting-state connectivity in a substantia nigra-related striato-thalamo-cortical network in a large sample of first-episode drug-naïve patients with schizophrenia. Schizophr Bull 2017;44:419–431.

[70] Luo F-F, Xu H, Zhang M, Wang Y. Abnormal regional spontaneous brain activity and its indirect effect on spasm ratings in patients with hemifacial spasm. Front Neurosci 2020;14:601088.

[71] Yang Y, Cui Q, Pang Y, Chen Y, Tang Q, Guo X, et al. Frequency-specific alteration of functional connectivity density in bipolar disorder depression. Prog Neuropsychopharmacol Biol Psychiatry 2021;104:110026.

[72] Zhang T, Zhao Z, Zhang C, Zhang J, Jin Z, Li L. Classification of early and late mild cognitive impairment using functional brain network of resting-state fMRI. Front Psychiatry 2019;10:572.

[73] Li Y, Yao H, Lin P, Zheng L, Li C, Zhou B, et al. Frequency-dependent altered functional connections of default mode network in Alzheimer's disease. Front Aging Neurosci 2017;9:259.

[74] Qian L, Zhang Y, Zheng L, Fu X, Liu W, Shang Y, et al. Frequency specific brain networks in Parkinson's disease and comorbid depression. Brain Imaging Behav 2017;11:224–239.

[75] Guan X, Guo T, Zeng Q, Wang J, Zhou C, Liu C, et al. Oscillation-specific nodal alterations in early to middle stages Parkinson's disease. Transl Neurodegener 2019;8:36.

[76] Xue SW, Li D, Weng XC, Northoff G, Li DW. Different neural manifestations of two slow frequency bands in resting functional magnetic resonance imaging: A systemic survey at regional, interregional, and network levels. Brain Connect 2014;4:242–255.

[77] Song X, Zhang Y, Liu Y. Frequency specificity of regional homogeneity in the resting-state human brain. PLoS One 2014;9:e86818.

[78] Kawashima R, Roland PE, O'Sullivan BT. Functional-anatomy of reaching and visuomotor learning-a positron emission tomography study. Cereb Cortex 1995;5:111–122.

[79] Kjaer TW, Nowak M, Lou HC. Reflective self-awareness and conscious states: PET evidence for a common midline parietofrontal core. Neuroimage 2002;17:1080–1086.

[80] Lundstrom BN, Petersson KM, Andersson J, Johansson M, Fransson P, Ingvar M. Isolating the retrieval of imagined pictures during episodic memory: Activation of the left precuneus and left prefrontal cortex. Neuroimage 2003;20:1934–1943.

[81] Cavanna AE, Trimble MR. The precuneus: A review of its functional anatomy and behavioural correlates. Brain 2006;129:564–583.

[82] Lou HC, Luber B, Crupain M, Keenan JP, Nowak M, Kjaer TW, et al. Parietal cortex and representation of the mental self. Proc Natl Acad Sci USA 2004;101:6827–6832.

[83] Vogt BA, Laureys S. Posterior cingulate, precuneal and retrosplenial cortices: Cytology and components of the neural network correlates of consciousness. In: Laureys S, editor. Boundaries of consciousness: Neurobiology and neuropathology. 2005. p. 205–217.

[84] Luo Z, Yin E, Zeng LL, Shen H, Su J, Peng L, et al. Frequency-specific segregation and integration of human cerebral cortex: An intrinsic functional atlas. iScience 2024; 27:109206.

[85] Han Y, Wang J, Zhao Z, Min B, Lu J, Li K, et al. Frequency-dependent changes in the amplitude of low-frequency fluctuations in amnestic mild cognitive impairment: A resting-state fMRI study. Neuroimage 2011;55:287–295.

[86] Han Y, Lui S, Kuang W, Lang Q, Zou L, Jia J. Anatomical and functional deficits in patients with amnestic mild cognitive impairment. PLoS One 2012;7:e28664.

[87] Jing L-L, Huang L-Y, Huang D-F, Niu J, Zhong Z. Amplitude of low frequency fluctuation at different frequency bands in early amnestic mild cognitive impairment: Results from ADNI. J Innov Opt Health Sci 2012;5:1150003.

[88] Liu X, Wang S, Zhang X, Wang Z, Tian X, He Y. Abnormal amplitude of low-frequency fluctuations of intrinsic brain activity in Alzheimer's disease. J Alzheimers Dis 2014;40:387–397.

[89] Li Y, Jing B, Liu H, Li Y, Gao X, Li Y, et al. Frequency-dependent changes in the amplitude of low-frequency fluctuations in mild cognitive impairment with mild depression. J Alzheimers Dis 2017;58:1175–1187.

[90] Veldsman M, Egorova N, Singh B, Mungas D, DeCarli C, Brodtmann A. Low-frequency oscillations in default mode subnetworks are associated with episodic memory impairments in Alzheimer's disease. Neurobiol Aging 2017;59:98–106.

[91] Tian Y, Ding Z, Tam KY, Wang Z, Zhang H, Zhao D, Zhao Y, Xu W, Zheng S. Specific frequency bands of amplitude low-frequency fluctuations in memory-related cognitive impairment: Predicting Alzheimer's disease. ADMET DMPK 2015;3:274–280.

[92] Wang S, Rao J, Yue Y, Xue C, Hu G, Qi W, et al. Altered frequency-dependent brain activation and white matter integrity associated with cognition in characterizing preclinical Alzheimer's disease stages. Front Hum Neurosci 2021;15:625232.

[93] Martino M, Magioncalda P, Huang Z, Conio B, Piaggio N, Duncan NW, et al. Contrasting variability patterns in the default mode and sensorimotor networks balance in bipolar depression and mania. Proc Natl Acad Sci USA 2016;113:4824–4829.

[94] Meda SA, Wang Z, Ivleva EI, Poudyal G, Keshavan MS, Tamminga CA, et al. Frequency-specific neural signatures of spontaneous low-frequency resting state fluctuations in psychosis: Evidence from bipolar-schizophrenia network on intermediate phenotypes (B-SNIP) consortium. Schizophr Bull 2015;41:1336–1348.

[95] Yang AC, Tsai SJ, Lin CP, Peng CK, Huang NE. Frequency and amplitude modulation of resting-state fMRI signals and their functional relevance in normal aging. Neurobiol Aging 2018;70:59–69.

[96] Liu S, Zhang C, Meng C, Wang R, Jiang P, Cai H, et al. Frequency-dependent genetic modulation of neuronal oscillations: A combined transcriptome and resting-state functional MRI study. Cereb Cortex 2022;32:5132–5144.

[97] Li W, Wang M, Wen W, Huang Y, Chen X, Fan W, et al. Neural dynamics during resting state: A functional magnetic resonance imaging exploration with reduction and visualization. Complexity 2018;4181649.

[98] Guillery RW, Sherman SM. Thalamic relay functions and their role in corticocortical communication: Generalizations from the visual system. Neuron 2002;33:163–175.

[99] Sherman SM, Guillery RW. The role of the thalamus in the flow of information to the cortex. Philos Trans R Soc B Biol Sci 2002;357:1695–1708.

[100] Kreitzer AC, Malenka RC. Striatal plasticity and basal ganglia circuit function. Neuron 2008;60:543–554.

[101] Silkis I. The cortico-basal ganglia-thalamocortical circuit with synaptic plasticity. I. Modification rules for excitatory and inhibitory synapses in the striatum. Biosystems 2000;57:187–196.

[102] Silkis I. The cortico-basal ganglia-thalamocortical circuit with synaptic plasticity. II. Mechanism of synergistic modulation of thalamic activity via the direct and indirect pathways through the basal ganglia. Biosystems 2001;59:7–14.

[103] Zhang J, Wei L, Hu X, Zhang Y, Zhou D, Li C, et al. Specific frequency band of amplitude low-frequency fluctuation predicts Parkinson's disease. Behav Brain Res 2013;252:18–23.

[104] Hou Y, Wu X, Hallett M, Chan P, Wu T. Frequency-dependent neural activity in Parkinson's disease. Hum Brain Mapp 2014;35:5815–5833.

[105] Zhang J, Gao Z, Hou Y, Zang Y, Feng T. Spontaneous neural activity in different frequency bands in Parkinson's disease: A fMRI study. Chin J Behav Med Brain Sci 2016;25:220–225.

[106] Wang J, Zhang J-R, Zang Y-F, Wu T. Consistent decreased activity in the putamen in Parkinson's disease: A meta-analysis and an independent validation of resting-state fMRI. GigaScience 2018;7:6.

[107] Hu H, Chen J, Huang H, Zhou C, Zhang S, Liu X, et al. Common and specific altered amplitude of low-frequency fluctuations in Parkinson's disease patients with and without freezing of gait in different frequency bands. Brain Imaging Behav 2020;14:857–868.

[108] Tian Z-Y, Qian L, Fang L, Peng X-H, Zhu X-H, Wu M, et al. Frequency-specific changes of resting brain activity in Parkinson's disease: A machine learning approach. Neuroscience 2020;436:170–183.

[109] Xue S, Wang X, Wang W, Liu J, Qiu J. Frequency-dependent alterations in regional homogeneity in major depression. Behav Brain Res 2016;306:13–19.

[110] Frühholz S, Trost W, Grandjean D, Belin P. Neural oscillations in human auditory cortex revealed by fast fMRI during auditory perception. NeuroImage 2020;207:116401.

[111] Zhou Z, Wang J-B, Zang Y-F, Pan G. Pair comparison between two within-group conditions of resting-state fMRI improves classification accuracy. Front Neurosci 2018;11:740.

[112] Zou H, Yang J. Multi-frequency dynamic weighted functional connectivity networks for schizophrenia diagnosis. Appl Magn Reson 2019;50:847–859.

[113] Küchenhoff S, Sorg C, Schneider SC, Kohl O, Müller HJ, Napiorkowski N, et al. Visual processing speed is linked to functional connectivity between right frontoparietal and visual networks. Eur J Neurosci 2021;53:3362–3377.

[114] Ries A, Chang C, Glim S, Meng C, Sorg C, Wohlschläger A. Grading of frequency spectral centroid across resting-state networks. Front Hum Neurosci 2018;12:436.

[115] Tadayonnejad R, Yang S, Kumar A, Ajilore O. Clinical, cognitive, and functional connectivity correlations of resting-state intrinsic brain activity alterations in unmedicated depression. J Affect Disord 2015;172:241–250.

[116] Bushara KO, Grafman J, Hallett M. Neural correlates of auditory-visual stimulus onset asynchrony detection. J Neurosci 2001;21:300–304.

[117] Bushara KO, Hanakawa T, Immisch I, Toma K, Kansaku K, Hallett M. Neural correlates of cross-modal binding. Nat Neurosci 2003;6:190–195.

[118] Uddin LQ, Supekar K, Menon V. Typical and atypical development of functional human brain networks: Insights from resting-state fMRI. Front Syst Neurosci 2010;4:21.

[119] Ma J, Lin Y, Hu C, Zhang J, Yi Y, Dai Z. Integrated and segregated frequency architecture of the human brain network. Brain Struct Funct 2021;226:335–350.

[120] Margulies DS, Ghosh SS, Goulas A, Falkiewicz M, Huntenburg JM, Langs G, et al. Situating the default-mode network along a principal gradient of macroscale cortical organization. Proc Natl Acad Sci U S A 2016;113:12574–12579.

[121] Gong Z-Q, Zuo X-N. Connectivity gradients in spontaneous brain activity at multiple frequency bands. Cereb Cortex 2023;33:9718–9728.

[122] Tomasi D, Wang GJ, Volkow ND. Energetic cost of brain functional connectivity. Proc Natl Acad Sci U S A 2013;110:13642–13647.

[123] Raichle ME, Gusnard DA. Appraising the brain's energy budget. Proc Natl Acad Sci U S A 2002;99:10237–10239.

[124] Tavor I, Jones OP, Mars RB, Smith SM, Behrens TE, Jbabdi S. Task-free MRI predicts individual differences in brain activity during task performance. Science 2016;352:216–220.

[125] Gong Z-Q, Zuo X-N. Cortical activations in cognitive task performance at multiple frequency bands. 2024;34:bhae489.

[126] Huntenburg JM, Bazin PL, Margulies DS. Large-scale gradients in human cortical organization. Trends Cogn Sci 2018;22:21–31.

[127] Huntenburg JM, Bazin PL, Goulas A, Tardif CL, Margulies DS. A systematic relationship between functional connectivity and intracortical myelin in the human cerebral cortex. Cereb Cortex 2017;27:1–17.

[128] Raut RV, Snyder AZ, Raichle ME. Hierarchical dynamics as a macroscopic organizing principle of the human brain. Proc Natl Acad Sci U S A 2020;117:20890–20897.

[129] Northoff G, Wainio-Theberge S, Evers K. Is temporo-spatial dynamics the "common currency" of brain and mind? In quest of "spatiotemporal neuroscience." Phys Life Rev 2020;33:34–54.

[130] Collin G, van den Heuvel MP. The ontogeny of the human connectome: Development and dynamic changes of brain connectivity across the life span. Neuroscientist 2013;19:616–628.

[131] Zuo X-N, He Y, Betzel RF, Colcombe S, Sporns O, Milham MP. Human connectomics across the life span. Trends Cogn Sci 2017;21:32–45.

[132] Hampel H, Vergallo A, Perry G, Lista S, Aguilar LF, Babiloni C, et al. The Alzheimer precision medicine initiative. J Alzheimers Dis 2019;68:1–24.

[133] Rutherford S, Fraza C, Dinga R, Kia SM, Wolfers T, Zabihi M, et al. Charting brain growth and aging at high spatial precision. Elife 2022;11:e72904.

[134] Park DC, Lautenschlager G, Hedden T, Davidson NS, Smith AD, Smith PK. Models of visuospatial and verbal memory across the adult life span. Psychol Aging 2002;17:299–320.

[135] Hertzog C, Dixon RA, Hultsch DF, MacDonald SW. Latent change models of adult cognition: Are changes in processing speed and working memory associated with changes in episodic memory? Psychol Aging 2003;18:755–769.

[136] Xing X-X. Globally aging cortical spontaneous activity revealed by multiple metrics and frequency bands using resting-state functional MRI. Front Aging Neurosci 2021;13:803436.

[137] Kessler RC, Amminger GP, Aguilar-Gaxiola G, Alonso J, Lee S, Ustun TB. Age of onset of mental disorders: A review of recent literature. Curr Opin Psychiatry 2007;20:359–364.

[138] Gui R, Chen T, Nie H. Classification of task-state fMRI data based on circle-EMD and machine learning. Comput Intell Neurosci 2020;2020:7691294.

[139] Chen L-Z, Holmes AJ, Zuo X-N, Dong Q. Neuroimaging brain growth charts: A road to mental health. Psychoradiology 2021;1:272–286.

[140] Toi PT, Jang HJ, Min K, Kim S-P, Lee S-K, Lee J, et al. In vivo direct imaging of neuronal activity at high temporospatial resolution. Science 2022;378:160–168.

[141] Finn ES. Is it time to put rest to rest? Trends Cogn Sci 2021;25:1021–1032.

[142] Huang Z. Temporospatial Nestedness in Consciousness: An Updated Perspective on the Temporospatial Theory of Consciousness. Entropy (Basel). 2023;25:1074.

[143] Bettinger JS, Friston KJ. Conceptual foundations of physiological regulation incorporating the free energy principle and self-organized criticality. Neurosci Biobehav Rev, 2023;155:105459.

[144] Bolt T, Nomi JS, Bzdok D, Salas JA, Chang C, Thomas Yeo BT, et al. A parsimonious description of global functional brain organization in three spatiotemporal patterns. Nat Neurosci 2022;25:1093-1103.

[145] Zhang D, Raichle ME. Disease and the brain's dark energy. Nat Rev Neurol 2010;6:15-28.